\documentclass[twocolumn,superscriptaddress,aps,prl,longbibliography,showpacs,preprintnumbers,amsmath,amssymb,floatfix]{revtex4-1}

\usepackage[latin9]{inputenc}
\usepackage{color}
\usepackage{dcolumn}
\usepackage{bm}
\usepackage{bbm}
\usepackage{braket}
\usepackage{mathrsfs}
\usepackage{amstext}

\usepackage{txfonts}

\usepackage{amsmath}
\usepackage{dsfont}
\usepackage{euscript}
\usepackage{amssymb}
\usepackage{graphicx}
\usepackage{esint}

\usepackage[unicode=true,
 bookmarks=false,
 breaklinks=false,pdfborder={0 0 1},backref=false,colorlinks=true,citecolor=blue]
 {hyperref}

\makeatletter
\@ifundefined{textcolor}{}
{%
 \definecolor{BLACK}{gray}{0}
 \definecolor{WHITE}{gray}{1}
 \definecolor{RED}{rgb}{1,0,0}
 \definecolor{GREEN}{rgb}{0,1,0}
 \definecolor{BLUE}{rgb}{0,0,1}
 \definecolor{CYAN}{cmyk}{1,0,0,0}
 \definecolor{MAGENTA}{cmyk}{0,1,0,0}
 \definecolor{YELLOW}{cmyk}{0,0,1,0}
}

\makeatletter
\renewcommand*\env@matrix[1][*\c@MaxMatrixCols c]{%
  \hskip -\arraycolsep
  \let\@ifnextchar\new@ifnextchar
  \array{#1}}
\makeatother

\newcommand{\eref}[1]{Eq.\,\eqref{#1}}
\newcommand{\fref}[1]{Fig.\,\ref{#1}}

\newcommand{\cref}[1]{Ref.\,\cite{#1}}

\newcommand{\oket}[1]{|#1\rangle\!\rangle}
\newcommand{\obra}[1]{\langle\!\langle #1|}

\newcommand{\mc}[1]{\mathcal{#1}}

\def\rhoab{\hat\rho_{\scriptscriptstyle\!\mathscr{A}\!\mathscr{B}}}

\makeatother

\begin{document}

\title{A solvable family of driven-dissipative many-body systems}

\author{Michael Foss-Feig}
\affiliation{United States Army Research Laboratory, Adelphi, Maryland 20783, USA}
\affiliation{Joint Quantum Institute, NIST/University of Maryland, College Park, Maryland 20742, USA}
\affiliation{Joint Center for Quantum Information and Computer Science, NIST/University of Maryland, College Park, Maryland 20742, USA}

\author{Jeremy T.\,Young}
\affiliation{Joint Quantum Institute, NIST/University of Maryland, College Park, Maryland 20742, USA}

\author{Victor V.\,Albert}
\affiliation{Yale Quantum Institute and Department of Physics, Yale University, New Haven, Connecticut 06520, USA}

\author{Alexey V.\,Gorshkov}
\affiliation{Joint Quantum Institute, NIST/University of Maryland, College Park, Maryland 20742, USA}
\affiliation{Joint Center for Quantum Information and Computer Science, NIST/University of Maryland, College Park, Maryland 20742, USA}

\author{Mohammad F.\,Maghrebi}
\affiliation{Department of Physics and Astronomy, Michigan State University, East Lansing, Michigan 48824, USA}
\affiliation{Joint Quantum Institute, NIST/University of Maryland, College Park, Maryland 20742, USA}
\affiliation{Joint Center for Quantum Information and Computer Science, NIST/University of Maryland, College Park, Maryland 20742, USA}

\begin{abstract}
Exactly solvable models have played an important role in establishing the sophisticated modern understanding of equilibrium many-body physics.  And conversely, the relative scarcity of solutions for \emph{non-equilibrium} models greatly limits our understanding of systems away from thermal equilibrium.  We study a family of non-equilibrium models, some of which can be viewed as dissipative analogues of the transverse-field Ising model, in that an effectively classical Hamiltonian is frustrated by dissipative processes that drive the system toward states that do not commute with the Hamiltonian.  Surprisingly, a broad and experimentally relevant subset of these models can be solved efficiently in any number of spatial dimensions.  We leverage these solutions to prove a no-go theorem on steady-state phase transitions in a many-body model that can be realized naturally with Rydberg atoms or trapped ions, and to compute the effects of decoherence on a canonical trapped-ion-based quantum computation architecture.
\end{abstract}


\maketitle

The understanding of equilibrium many-body physics in both classical and quantum systems has relied heavily on exact solutions of highly simplified models. For example, solutions of the classical and quantum (transverse-field) Ising models have elucidated the structure of classical and quantum phase transitions, respectively, and the deep connections between them \cite{RevModPhys.69.315,sachdev_book}. For quantum systems that are \emph{not} in thermal equilibrium such solutions are comparatively scarce, though important progress has been made in numerous specialized models that either: (a) are one dimensional and isolated \cite{1742.5468.2005.04.P04010,PhysRevA.78.010306}, (b) impose nontrivial dissipation on otherwise free bosons or fermions \cite{1367.2630.10.4.043026,PhysRevA.87.012108,Huffman2016309,Znidaric:2010if,Znidaric:2011dp,Eisler:2011fb,Zunkovic:2014fw,Guo_bosons,Guo_fermions}, (c) impose highly fine-tuned dissipation \cite{Zunkovic:2014fw}, or (d) are coupled to an environment only at a boundary \cite{PhysRevLett.101.105701,PhysRevLett.106.217206,PhysRevLett.107.137201,1751-8121-48-37-373001}. On the other hand, interacting quantum spin systems that are dissipative in the bulk and driven so that they do not thermalize are expected to exhibit a variety of rich behaviors not found in their equilibrium counterparts, including unusual multi-critical points \cite{Lee13}, critical exponents not found in or near equilibrium \cite{0034.4885.79.9.096001}, and the existence of zero-entropy entangled steady states \cite{PhysRevA.78.042307,nphys1073,nphys1342,Pastawski_2014}.  And, naturally, such systems play an important role in the theory of quantum computation in the presence of decoherence.  While recent experimental advances have enabled the controlled study of such physics in systems ranging from exciton-polariton fluids \cite{Deng199,Kasprazak2006,PhysRevLett.96.230602,Byrnes2014,ncomms11887}, to trapped ions \cite{Bohnet1297,10.1038/nphys2630}, to Rydberg gases \cite{10.1038/nature11361,Carr13,PhysRevLett.113.023006}, the minimal microscopic models expected to capture the essential qualitative physics---many-body quantum master equations \cite{0034.4885.79.9.096001}---continue to pose severe challenges to existing theoretical techniques.

\begin{figure}[!t!]
\includegraphics[width=1.0\columnwidth]{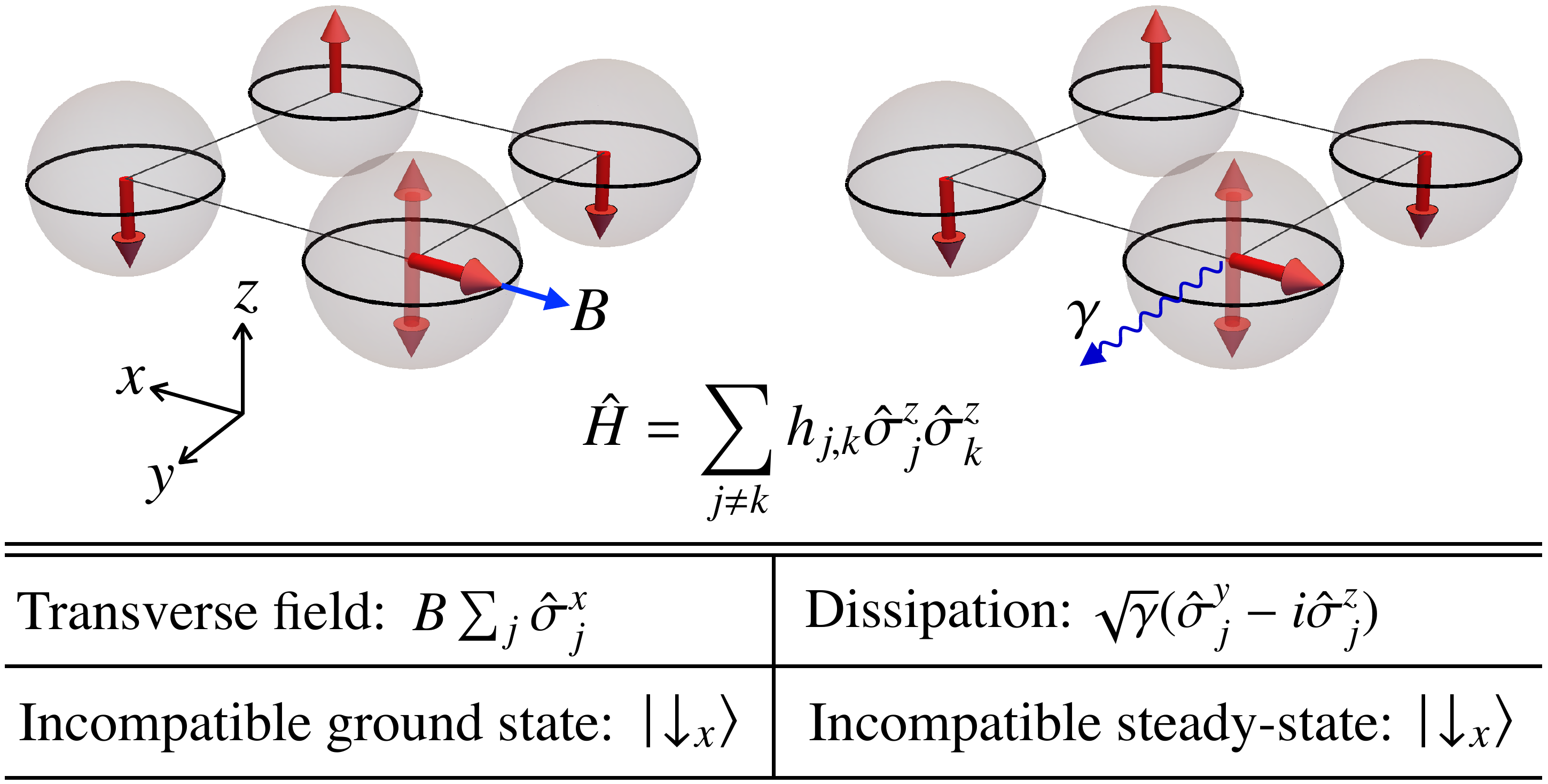}
\caption{(Color online) In the transverse field Ising model (left panel) ordering with respect to a ``classical'' Hamiltonian $\hat{H}$ is frustrated, even at $T=0$, because the ground state of the transverse field (expressed as a density matrix) does not commute with $\hat{H}$.  The models solved here generalize this scenario to the case where order is frustrated by the inclusion of a \emph{dissipative process} whose \emph{steady state} does not commute with $\hat{H}$.  In the example shown (right panel), Markovian dissipation implemented by jump operators $\hat{\sigma}^y_j-i\hat\sigma^z_j$ drives each spin towards the ground-state of a transverse field.}
\label{fig:schematic}
\end{figure}

In this manuscript we investigate a broad and experimentally relevant class of driven-dissipative many-body spin models, some of which can be viewed as dissipative analogues of the transverse-field Ising model (TFIM), and show that they can be solved efficiently.  The TFIM captures a characteristic feature of low-temperature quantum systems more generally: even at zero temperature, the ordering associated with a classical Hamiltonian (the Ising model) can be frustrated by the persistence of quantum fluctuations (induced by the transverse field, left panel of \fref{fig:schematic}).  In particular, the transverse field favors a zero-temperature density matrix that does not commute with the Ising Hamiltonian. As a result, energy minimization forces the system to develop quantum correlations even between arbitrarily distant spins, generically prohibiting exact calculations of its ground-state properties (except in certain very well-known circumstances, e.g.\ nearest-neighbor interactions in 1D).  The models we consider realize a natural non-equilibrium analogue of this scenario, in which a classical Hamiltonian is frustrated by the presence of \emph{dissipative} fluctuations (right panel, \fref{fig:schematic}). Strikingly, we find that the inclusion of a broad class of dissipative processes favoring a \emph{steady-state} density matrix that does not commute with the Hamiltonian is relatively benign, even in more than one spatial dimension.  In particular, correlations remain dynamically localized whenever the Hamiltonian is of finite range, enabling the time-dependence of finite-order correlation functions to be efficiently obtained (even for an infinite system) by solving a finite-dimensional system of equations. This structure exists, for example, even when the dissipation alone drives the system towards the ground state of a transverse field.  It also exists when the dissipation is explicitly derived from a temporally fluctuating transverse field; evidently, even though the Ising model is not (in general) expected to be solvable in the presence of a \emph{particular} (static or time-varying) transverse field, the dynamics of the transverse-field Ising model averaged over \emph{all} such fields can be computed exactly.

To illustrate the utility and generality of these solutions, we use them to prove that an experimentally relevant subset of the models considered possess a finite dissipative gap, and therefore cannot undergo dissipative phase transitions.  We then apply this result to a model of interacting Rydberg atoms studied in Refs.\ \cite{PhysRevB.93.014307,Weimer2016}, thereby confirming the structure of the phase diagram inferred from recently developed approximate techniques \cite{PhysRevLett.114.040402}.  From a more applied perspective, we emphasize that Ising-like Hamiltonians, despite being ``classical'', induce truly quantum dynamics and play a central role in the production of states sought for quantum information processing and quantum metrology \cite{PhysRevLett.86.5188,PhysRevA.47.5138,PhysRevLett.82.1835}. Their solvability in the presence of dissipation thus affords numerous exciting opportunities to investigate the effect of decoherence on the generation of useful entangled states.  As an illustrative example, we compute the effects of a fluctuating transverse field (dephasing) on Ising dynamics generated by M\o{}lmer-S\o{}rensen gates \cite{PhysRevLett.82.1835,1464.4266.7.10.025,10.1038/nature09071}, which can impose important technical limitations in trapped-ion based approaches to quantum computation \cite{PhysRevLett.117.060505}.


\emph{Dissipative models.---}The models we consider can be constructed in close analogy to the transverse-field Ising model.  There, one starts with a many-body Hamiltonian that is ``classical'' in the sense that it can be diagonalized by a choice of local basis.  Thus its eigenstates have well defined local properties, and in this sense there are no quantum fluctuations \cite{FN1}.  To be concrete we restrict our focus to the spin-$\frac{1}{2}$ Ising model,
\begin{align}
\label{eq:Hamiltonian}
\hat{H}=\sum_{j\neq k}h_{j,k}\hat{\sigma}^z_j\hat{\sigma}^z_k.
\end{align}
However, the scope of our results is significantly broader, and includes generic Hamiltonians that can be diagonalized by a choice of local basis or, with some additional constraints, commuting Hamiltonians that are not diagonalized by a choice of local basis (e.g.\ the toric code \cite{Kitaev20032}).  Extensions along these lines will be discussed at the end of the manuscript.  In equilibrium, a natural question to ask is how one can modify $\hat{H}$ such that, in the low temperature limit where thermal fluctuations vanish, quantum fluctuations remain.  One simple strategy is to add single-body terms to the Hamiltonian that are not diagonalized by the eigenstates of $\hat{H}$; the usual culprit is a transverse field, resulting in an additional term $B\sum_{j}\hat{\sigma}_j^x$.  Even at zero temperature, i.e.\ in the quantum ground state, such a term induces fluctuations between the classical eigenstates of the interaction part of the Hamiltonian.

In a driven-dissipative setting, we are no longer interested in properties of the ground state, which control the low-temperature equilibrium physics, but rather the properties of the steady state, which control the long-time, non-equilibrium physics.  A natural dissipative generalization of the procedure used above to frustrate the zero-temperature ordering associated with $\hat{H}$ is to introduce single-body dissipative processes that drive the system toward a steady state that does not commute with $\hat{H}$. Assuming that dissipation can be treated in the Born-Markov approximation, the dynamics of an open quantum system with Hamiltonian $\hat{H}$ is governed by a Markovian master equation of the form
\begin{align}
\dot{\hat{\rho}}&=\mathcal{L}(\hat{\rho})\equiv -i[\hat{H},\hat{\rho}]+\mathcal{D(\hat{\rho})},\nonumber\\
\label{eq:dissipator}
\mathcal{D}(\star)&\equiv\sum_{j,\alpha}\frac{\gamma_{j\alpha}}{2}\big(2\hat{J}^{\phantom\dagger}_{j\alpha}\star \hat{J}^{\dagger}_{j\alpha}-\{\hat{J}^{\dagger}_{j\alpha}\hat{J}^{\phantom\dagger}_{j\alpha},\star\}\big).
\end{align}
The dissipation is induced by jump operators $\hat{J}_{j\alpha}$, each of which we assume to be supported on a single site $j$. The index $\alpha$ may take on multiple values in order to describe multiple dissipative channels on a given site, though in what follows we will generally consider only one jump operator on each site, dropping the index $\alpha$.  Consider the dissipative dynamics in the absence of the Hamiltonian, described by $\dot{\hat{\rho}}=\mathcal{D}(\hat{\rho})$.  The steady-state solution of the purely dissipative dynamics is determined implicitly by solving $\mathcal{D}(\hat{\rho}_{\rm dis})=0$. If $\hat{\rho}_{\rm dis}$ commutes with the Hamiltonian, then it is automatically also a solution of $\mathcal{L}(\hat{\rho})=0$ and thus is a proper steady state of the complete dynamics including both coherent evolution and dissipation.  If it does not, then in close analogy to the ground state of the transverse-field Ising model, we expect the steady state to possess fluctuations in the sense of admixing of classical states.

\begin{figure}[!t]
\includegraphics[width=0.94\columnwidth]{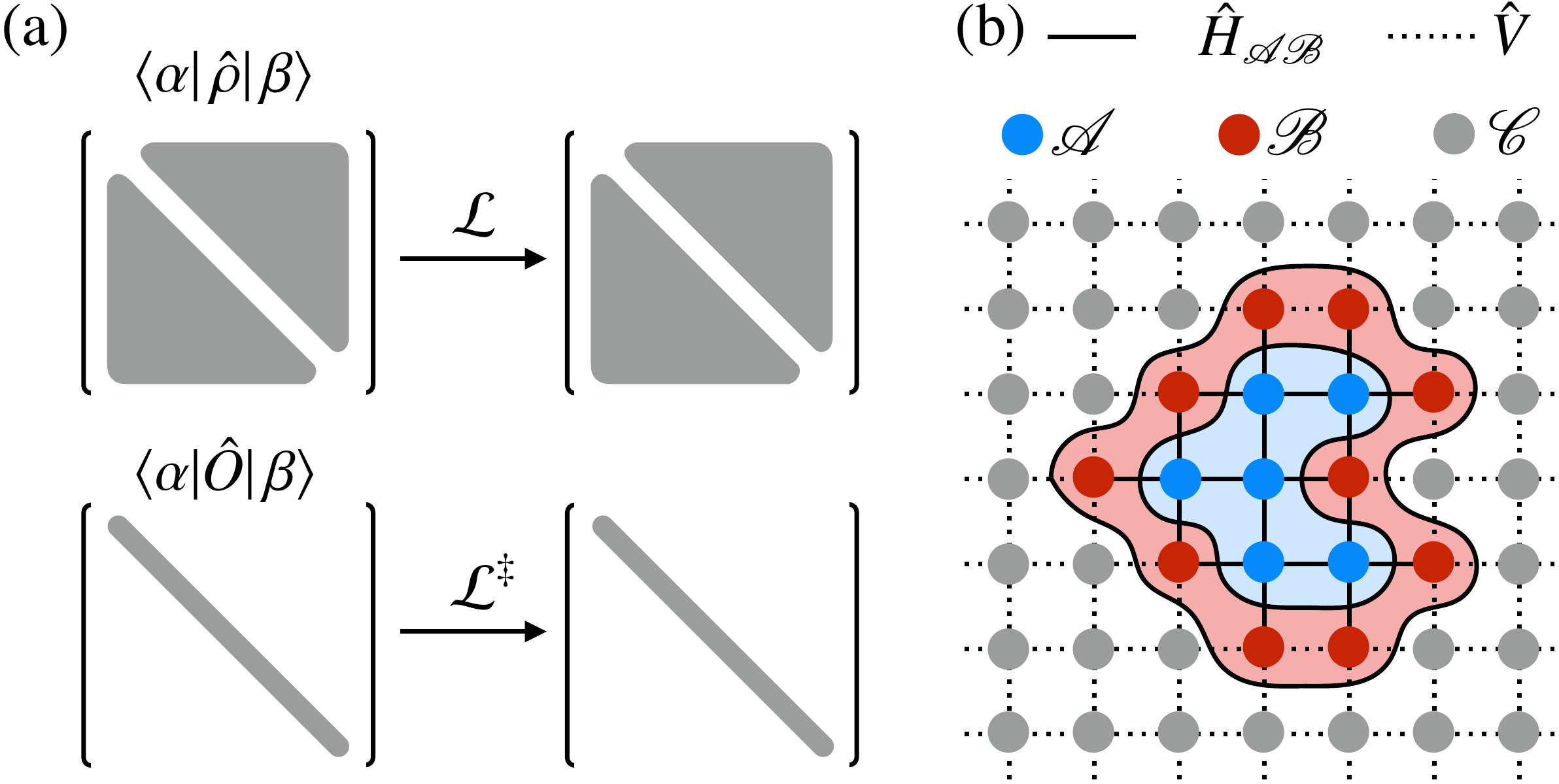}
\caption{(Color online) (a) Schematic representation of \eref{eq:constraint_2}: In the $z$ basis, $\mathcal{L}$ does not map coherences to populations.  Equivalently, in the Heisenberg picture, $\mathcal{L}^{\dagger}$ does not cause diagonal operators to develop coherences. (b)  When the conditions described in (a) are met, the dynamics of any observable supported on a subsystem $\mathscr{A}$ can be computed by identifying the nearest neighbors of the set $\mathscr{A}$, denoted $\mathscr{B}$, tracing out the part of the density matrix supported on $\mathscr{C}$, and then solving the master equation projected into  the remaining finite-dimensional system.}
\label{fig:lattice}
\end{figure}

Surprisingly, a large class of dissipators that frustrate the classical Hamiltonian, i.e. for which $[\hat{\rho}_{\rm dis},\hat{H}]\neq 0$, nevertheless admit exact solutions for the dynamics of observables.  In particular, suppose that \cite{FN2}
\begin{align}
\label{eq:constraint}
{\rm Tr}\big(\hat{\sigma}^z_j\mathcal{D}(\hat{\sigma}^{\pm}_j)\big)=0~~~({\rm for~all}~j).
\end{align}
For $\hat{H}$ of the form in \eref{eq:Hamiltonian}, we also have ${\rm Tr}(\hat{\sigma}^z_j[\hat{H},\hat{\sigma}^{\pm}_j])=0$, and therefore whenever the jump operators are chosen to satisfy \eref{eq:constraint}, the complete Liouvillian also obeys
\begin{align}
\label{eq:constraint_2}
{\rm Tr}\big(\hat{\sigma}^z_j\mathcal{L}(\hat{\sigma}^{\pm}_j)\big)=0~~~({\rm for~all}~j).
\end{align}
Taken together with the equality ${\rm Tr}(\hat{\mathds{1}}_j\mathcal{L}(\hat{\sigma}^{\pm}_j))=0$, which follows trivially from the definition of $\mathcal{L}$ (and ensures the conservation of probability) regardless of the form of the jump operators, \eref{eq:constraint_2} can be understood colloquially as a statement that $\mathcal{L}$ does not map coherences onto populations (\fref{fig:lattice}a).  In what follows, we will show that finite-range Hamiltonians of the form in \eref{eq:Hamiltonian}, subjected to dissipation obeying \eref{eq:constraint}, produce correlations that are localized in space and can therefore be solved efficiently in the thermodynamic limit.

There are many natural jump operators satisfying \eref{eq:constraint}.  For example, dephasing, spontaneous emission, and incoherent pumping along the $z$ axis all do, and the models studied here therefore subsume finite-range versions of the models studied in \cite{PhysRevA.87.042101,1367.2630.15.11.113008} (and realized experimentally in \cite{Bohnet1297}) as special cases.  However, all the examples of dissipation just mentioned lead to steady states $\hat{\rho}_{\rm dis}$ that commute with $\hat{H}$, and thus the steady states in the presence of $\hat H$ remain trivial.  Examples of jump operators satisfying \eref{eq:constraint} but producing steady states that \emph{do not} commute with $\hat H$ include dephasing along any direction in the $xy$ plane; $\hat J=\cos(\theta)\hat\sigma^x+\sin(\theta)\hat\sigma^y$, or spontaneous emission along any axis in the $xy$ plane; $\hat J=\hat{\sigma}^z+i(\cos(\theta)\hat\sigma^x+\sin(\theta)\hat\sigma^y)$. Note that in the latter example, $\hat{\rho}_{\rm dis}$ is the ground state of a transverse field.

\emph{Localization of correlations}.---The time-dependent expectation value of an arbitrary operator $\hat{O}$, initially supported on a set of sites $\mathscr{A}$, can be written as
\begin{align}
\begin{matrix}
{\text{\small Schr\"odinger picture:}} & {\text{\small Heisenberg picture:}} \\
O(t)={\rm Tr}\big(\hat O \exp(t\mathcal{L})\,\hat\rho_{0}\big),~ & O(t)={\rm Tr}\big(\hat\rho_{0} \exp(t\mathcal{L}^{\ddagger})\,\hat O\big).
\end{matrix}
\end{align}
The Heisenberg picture expression utilizes the adjoint Liouvillian $\mathcal{L}^{\ddagger}(\hat O)=i[\hat H,\hat O]+\mathcal{D}^{\ddagger}(\hat O)$, with the adjoint dissipator given by $\mathcal{D}^{\ddagger}(\star)=\sum_{j,\alpha}\frac{\gamma_{j\alpha}}{2}\big(2\hat J^{\dagger}_{j\alpha}\star \hat J^{\phantom\dagger}_{j\alpha}-\{\hat J^{\dagger}_{j\alpha}\hat J^{\phantom\dagger}_{j\alpha},\star\}\big)$. The operator $\hat O$ could be, for example, a product of two spin operators on different sites, in which case $O(t)$ is an equal-time correlation function (generalizations to unequal time correlation functions are also possible using the quantum regression formula). Equation \eqref{eq:constraint_2} can be recast in terms of the Heisenberg-picture Liouvillian as ${\rm Tr}[\hat \sigma^{\pm}_j\mathcal{L}^{\ddagger}(\hat \sigma^z_j)]=0$; thus, $\mathcal{L}^{\ddagger}$ does not map populations onto coherences (\fref{fig:lattice}a).  This condition on $\mathcal{L}^{\ddagger}$ can be restated in the following useful way: Referring to an operator as ``diagonal on the set of sites $\mathscr{S}$'' if it commutes with all operators $\hat{\sigma}_j^z$ for $j\in \mathscr{S}$, one can show that the set of operators that are diagonal on $\mathscr{S}$ is closed under the action of $\mathcal{L}^{\ddagger}$.  With this structure in mind, we refer to $\mathcal{L}^{\ddagger}$ as \emph{diagonality preserving} (in the $z$ basis), a property which will play a key role in the solution of \eref{eq:dissipator}.

To compute $O(t)$, it is useful to define the set of sites that are nearest neighbors of $\mathscr{A}$, denoted $\mathscr{B}$, as all sites outside of $\mathscr{A}$ for which a term in the Hamiltonian has simultaneous support on $\mathscr{A}$ and $\mathscr{B}$.  The complement of $\mathscr{A}\bigcup\mathscr{B}$, denoted $\mathscr{C}$, is therefore not directly connected to $\mathscr{A}$ by terms in the Hamiltonian (see \fref{fig:lattice}b).  We then decompose the Hamiltonian as $\hat H=\hat H_{\mathscr{A}\!\mathscr{B}}+\hat{H}_V$, where $\hat H_{\mathscr{A\!B}}$ contains all terms in $\hat{H}$ that have support on $\mathscr{A}$, and $\hat{H}_V$ contains all terms that do not (note that, by the definition of $\mathscr{B}$, $\hat H_{\mathscr{A\!B}}$ is supported on $\mathscr{A}\bigcup\mathscr{B}$, while $\hat{H}_V$ is supported on $\mathscr{B}\bigcup\mathscr{C}$).  Similarly, we decompose the Heisenberg-picture dissipator as $\mathcal{D}^{\ddagger}=\mathcal{D}^{\ddagger}_{\mathscr{A\!B}}+\mathcal{D}^{\ddagger}_{V}$, where $\mathcal{D}^{\ddagger}_{\mathscr{A\!B}}$ only contains jump operators supported on $\mathscr{A}\bigcup\mathscr{B}$ and $\mathcal{D}_{V}^{\ddagger}$ only contains jump operators supported on $\mathscr{C}$.  Writing $\mathcal{L}^{\ddagger}=\mathcal{L}^{\ddagger}_\mathscr{A\!B}+\mathcal{L}^{\ddagger}_V$, with $\mathcal{L}^{\ddagger}_{\mathscr{A\!B}}(\star)=i[\hat H_{\mathscr{A\!B}},\star]+\mathcal{D}_{\mathscr{A\!B}}^{\ddagger}(\star)$ and $\mathcal{L}^{\ddagger}_V(\star)=i[\hat{H}_V,\star]+\mathcal{D}^{\ddagger}_{V}(\star)$, we can write $O(t)$ as
\begin{align}
\label{eq:O_of_t}
O(t)={\rm Tr}_{\mathscr{A}\bigcup\mathscr{B}}\big({\rm Tr}_{\mathscr{C}}\big(\hat\rho_0 \sum_{n=0}^{\infty}\frac{t^n}{n!}\hat{O}_n\big)\big),
\end{align}
where $\hat{O}_n=(\mathcal{L}^{\ddagger}_\mathscr{A\!B}+\mathcal{L}^{\ddagger}_V)^n\hat O$. The structure of \eref{eq:O_of_t} can be greatly simplified by induction.  In particular, suppose that $\hat{O}_n$ satisfies the following two conditions:
\begin{quote}
(A) $\hat{O}_n$ is supported on $\mathscr{A}\bigcup\mathscr{B}$.
(B) $\hat{O}_n$ is diagonal on $\mathscr{B}$.
\end{quote}
Note that both conditions are satisfied trivially for $\hat{O}_0=\hat{O}$, since we have assumed that $\hat O$ is fully supported on $\mathscr{A}$.  We will show that the operator $\hat O_{n+1}=(\mathcal{L}^{\ddagger}_\mathscr{A\!B}+\mathcal{L}^{\ddagger}_{V})\hat O_n$ also satisfies these conditions, thereby proving that all $\hat O_n$ do.

Toward this end, we first show that $\mathcal{L}^{\ddagger}_V(\hat{O}_n)=i[\hat{H}_V,\hat{O}_n]+\mathcal{D}^{\ddagger}_V(\hat{O}_n)$ vanishes by the assumptions.  The first term vanishes because $\hat{H}_V$ is diagonal and supported on $\mathscr{B}\bigcup\mathscr{C}$, while $\hat{O}_n$ is diagonal on $\mathscr{B}\bigcup \mathscr{C}$ by conditions (A,B), giving $[\hat H_V,\hat O_n]=0$.  The second term vanishes because $\mathcal{D}^{\ddagger}_V$ only contains jump operators supported on $\mathscr{C}$, implying (together with condition A) that $\mathcal{D}^{\ddagger}_{V}(\hat O_n)=0$. Thus $\mathcal{L}^{\ddagger}_V( \hat O_n)=0$ as claimed, giving $\hat O_{n+1}=\mathcal{L}^{\ddagger}_{\mathscr{A\!B}}(\hat O_n)$. Now we note that, because $\hat H_{\mathscr{A\!B}}$ and the jump operators in $\mathcal{D}^{\ddagger}_{\mathscr{A\!B}}$ are supported on $\mathscr{A}\bigcup\mathscr{B}$, $\mathcal{L}^{\ddagger}_{\mathscr{A\!B}}(\hat O_n)$ must also be supported on $\mathscr{A}\bigcup\mathscr{B}$, and $\hat O_{n+1}$ therefore satisfies condition (A).  Furthermore, because $\mathcal{L}^{\ddagger}_{\mathscr{A}\!\mathscr{B}}$ is diagonality preserving, $\hat O_{n+1}$ continues to satisfy condition (B).  Therefore, by induction, we deduce that $\hat O_n=(\mathcal{L}^{\ddagger}_{\mathscr{A\!B}})^n\hat O$, and is supported on $\mathscr{A}\bigcup\mathscr{B}$.  Plugging this result into \eref{eq:O_of_t} and using the isolated support of $\hat O_n$ to carry out the trace over $\mathscr{C}$, we obtain
\begin{align}
O(t)={\rm Tr}_{\mathscr{A}\bigcup\mathscr{B}}\big(\rhoab\sum_{n=0}^{\infty}\frac{t^n}{n!}(\mathcal{L}^{\ddagger}_{\mathscr{A\!B}})^n\hat O\big),
\end{align}
where $\rhoab={\rm Tr}_{\mathscr{C}}(\hat \rho_0)$ is the reduced density matrix of the initial state upon tracing out all sites in $\mathscr{C}$.  Returning to the Schr\"odinger picture we find our primary result,
\begin{align}
\label{eq:final_ev}
O(t)={\rm Tr}_{\mathscr{A}\bigcup\mathscr{B}}\big(\hat O\exp(t\mathcal{L}_{\mathscr{A\!B}})\rhoab\big).
\end{align}
Equation \eqref{eq:final_ev} can be read as follows: The time-dependent expectation value of $\hat O$ can be computed by first calculating the reduced density matrix on $\mathscr{A}\bigcup\mathscr{B}$, and then evolving it using a Hamiltonian and jump operators supported only on $\mathscr{A}\bigcup\mathscr{B}$.  It follows immediately that correlations are localized for finite-ranged Hamiltonians.  For example, consider a correlation function $\langle\hat{O}\rangle=\langle\hat\sigma^{\mu}_j\hat\sigma^{\nu}_k\rangle$ ($\mu,\nu\in\{x,y,z\}$), and suppose the system starts in a product state.  If the Hamiltonian is of finite range $r$, and $j$ and $k$ are separated by a distance $d_{jk}>2r$ (so that sites $j$ and $k$ do not share any neighbors, and $\mathscr{A}\bigcup\mathscr{B}$ decomposes into two disjoint sets), then it follows from \eref{eq:final_ev} that $\langle\hat\sigma^{\mu}_j\hat\sigma^{\nu}_k\rangle=\langle\hat\sigma^{\mu}_j\rangle\langle\hat\sigma^{\nu}_k\rangle$ at all times \cite{FN3}.

\emph{Applications}.---Just as many-body ground states can suddenly change in character as a parameter in the Hamiltonian is continuously adjusted, signaling a quantum phase transition, the steady state of a master equation can exhibit a sudden change when a parameter in the Liouvillian is adjusted continuously, signaling a dissipative phase transition.  While a quantum phase transition is associated with the closing of an energy gap in the spectrum of the Hamiltonian, a dissipative phase transition is associated with the closing of a \emph{dissipative gap} in the spectrum of the Liouvillian. Consider the dissipative transverse-field Ising model studied in Refs.\,\cite{PhysRevB.93.014307,Weimer2016},
\begin{align}
\label{eq:tfim}
\hat{H}_{\rm TFIM}=J\sum_{\langle j,k\rangle}\hat{\sigma}^x_j\hat{\sigma}_k^x+\Delta\sum_{j}\hat{\sigma}^z_j,~~~~~\hat{J}_{j}=\hat{\sigma}^{-}_j,
\end{align}
in which fluctuations due to both dissipation (strength $\gamma$) \emph{and} a transverse field (strength $\Delta$) are considered; note that the state which minimizes the energy with respect to the transverse field (for $\Delta>0$) is identical to the steady state of the dissipation alone.  In \eref{eq:tfim} we have chosen to make the Hamiltonian diagonal along the $x$ direction for consistency with Ref.\ \cite{Weimer2016}.  However, for $\Delta=0$ and upon making a rotation of the coordinate system that takes the Ising part of $\hat{H}_{\rm TFIM}$ into the form of \eref{eq:Hamiltonian}, it can easily be verified that this model satisfies \eref{eq:constraint}.  In Ref.\,\cite{Weimer2016}, a careful analysis of the variational techniques developed in \cite{PhysRevLett.114.040402} suggests that the system is disordered for any value of $\gamma$ at $\Delta=0$ and in high dimensions.  Here, we can make this conclusion rigorous in \emph{any} dimension: In the supplement, we prove that this model at $\Delta=0$ [and also a more general subclass of the models obeying \eref{eq:constraint}] has a finite dissipative gap $\geq\gamma$.  Since the model is disordered at $\gamma=\infty$ and a gap exists at $\Delta=0$ for any $\gamma>0$, it must be disordered along the entire $\Delta=0$ axis.  In light of the Liouvillian stability results proved in Ref.\,\cite{Cubitt2015}, the existence of a dissipative gap at $\Delta=0$ may actually imply something much stronger: the gap persists for small $\Delta>0$, ruling out a phase transition for $\Delta\lesssim\gamma$ and implying that the system is disordered for sufficiently small $\Delta$ at any finite $\gamma$.

\begin{figure}[!t]
\includegraphics[width=0.74\columnwidth]{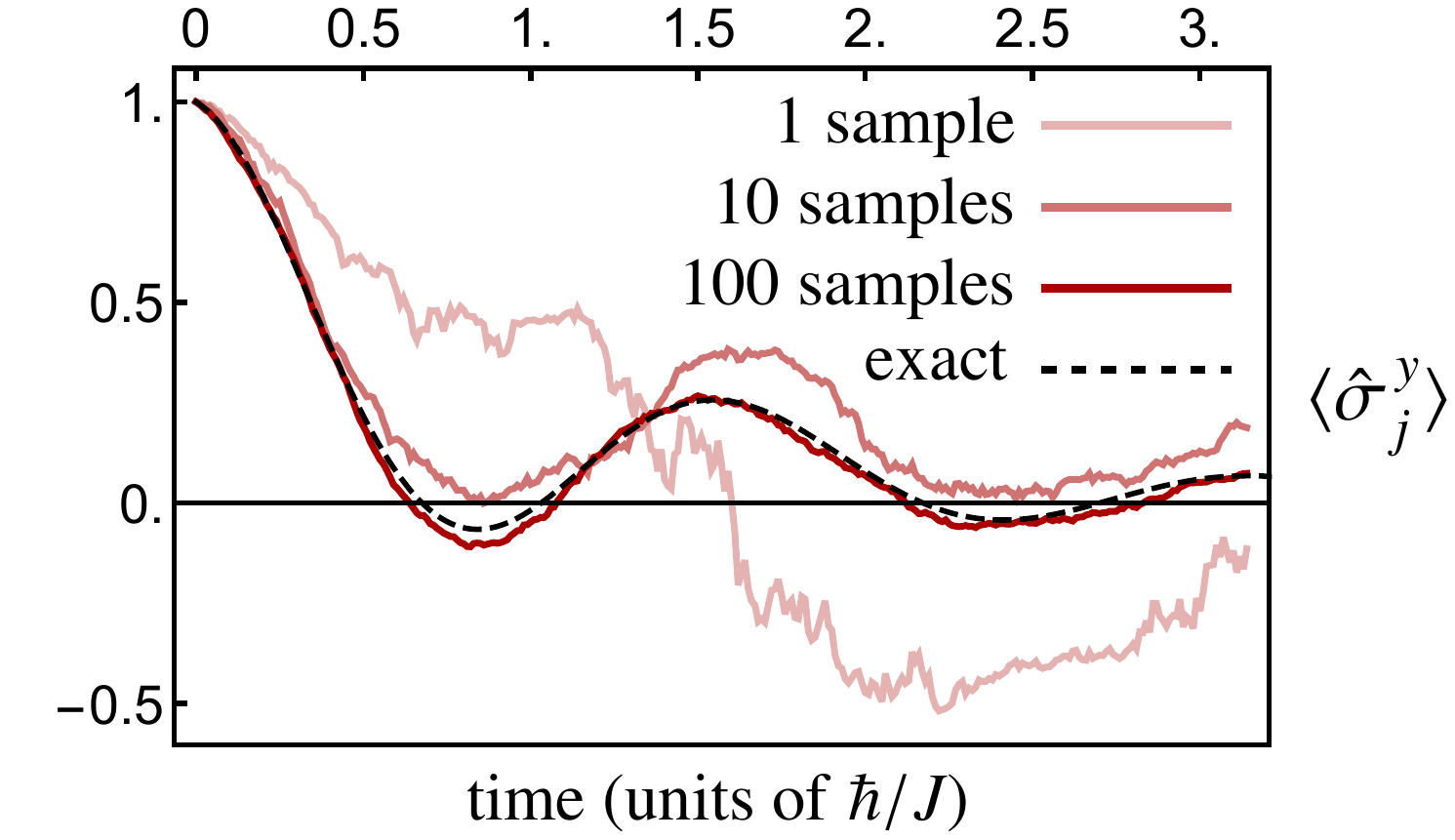}
\caption{(Color online) (a) Quench in the fluctuating-transverse-field Ising model starting with all spins polarized along $+y$ (10 spins in 1D with periodic boundary conditions, and $\gamma=J/4$).}
\label{fig:applications}
\end{figure}

The solutions developed here can also be used to calculate dynamics of the fluctuating-transverse-field Ising model,
\begin{align}
\label{eq:fTFIM}
\hat{H}_{\rm fTFIM}(t)=J\sum_{\langle j,k\rangle}\hat{\sigma}^x_j\hat{\sigma}_k^x+\sum_{j}\Delta_j(t)\,\hat{\sigma}^z_j.
\end{align}
Here, the transverse fields $\Delta_j(t)$ are Gaussian random variables with white-noise spectrum $\overline{\Delta_j(t_1)\Delta_k(t_2)}=\gamma \delta_{j,k}\delta(t_1-t_2)$.  This model---interpreted such that observables are obtained by averaging dynamics induced by $\hat H_{\rm fTFIM}(t)$ over realizations of the fluctuating transverse field---can be viewed as a continuous unraveling of the master equation describing dephasing dynamics in the Ising model \cite{Gisin1992,barchielli_book},
\begin{align}
\label{eq:ising_dephase}
\dot{\hat{\rho}}=-i[J\sum_{\langle j,k\rangle}\hat{\sigma}^x_j\hat{\sigma}_k^x,\hat\rho]+\gamma\sum_{j}(\hat\sigma^z_j\,\hat\rho\,\hat\sigma_j^z-\hat\rho).
\end{align}
[See also Refs.\ \cite{PhysRevA.64.052110,Chenu2016} for a constructive approach to \eref{eq:ising_dephase} starting from \eref{eq:fTFIM}]. A master equation  of this form arises naturally in ion trap experiments, where it captures the effects of qubit dephasing during Ising dynamics induced by M\o{}lmer-S\o{}rensen gates \cite{PhysRevLett.82.1835,1464.4266.7.10.025,10.1038/nature09071,PhysRevLett.117.060505}, and hence it plays a prominent role in the description of decoherence effects on trapped-ion based approaches to quantum computation. Naively one would not expect there to be an efficient solution to \eref{eq:ising_dephase}, given that its unraveling [either continuous or discrete, with the continuous case leading back to \eref{eq:fTFIM}] requires one to solve the Ising model in the presence of a time-dependent transverse field.  Nevertheless, after a rotation by $\pi/2$ about the $y$ axis \eref{eq:ising_dephase} falls into the class of solvable models studied here, and observables can be computed efficiently.  In \fref{fig:applications}(a), we show simulations of the stochastic Schr\"odinger equation associated with $\hat H_{\rm fTFIM}$; while these simulations are computationally very expensive, and not feasible for more than $\sim 10$ spins, the average over sufficiently many trajectories (lighter to darker red lines indicating more averaging) can be seen to converge to the exact solution (black-dashed line).

\emph{Outlook}.--- We have identified an exactly solvable class of driven-dissipative quantum spin models in which dissipation frustrates the ordering of a classical Hamiltonian. The underlying algebraic structure that enables an exact solution, ensured by \eref{eq:constraint}, can also be identified in a much more general class of Hamiltonians that are solvable in the absence of dissipation.  For example, the results can easily be generalized to systems with arbitrary finite-dimensional Hilbert spaces on each site and time-dependent Hamiltonians.  Long-range interactions do spoil some aspects of the solution, but the absence of dissipative phase transitions can still be proven.  Hamiltonians that are not diagonalized by a local choice of basis but can still be written as a sum of local commuting terms, such as the toric code \cite{Kitaev20032,Pastawski_2014}, are also amenable to similar solution techniques, and will be the topic of future research.

\begin{acknowledgments}
We acknowledge Hendrik Weimer and Vincent Overbeck for helpful discussions. M.F.M., J.T.Y., and A.V.G. acknowledge support by ARL CDQI, ARO MURI, NSF QIS, ARO, NSF PFC at JQI, and AFOSR. MM acknowledges start-up funding from Michigan State University.
\end{acknowledgments}


%


\newpage

\begin{widetext}

\appendix

\renewcommand{\thesection}{S\arabic{section}} 
\renewcommand{\theequation}{S\arabic{equation}}
\renewcommand{\thefigure}{S\arabic{figure}}
\setcounter{equation}{0}
\setcounter{figure}{0}

\section*{Supplemental Material}

This supplemental material contains a proof that, under conditions specified in the manuscript and reiterated below, the Liouvillians we consider possesses a finite dissipative gap. In particular, the dissipative gap of the Liouvillian $\mathcal{L}$ is bounded below by that of the dissipator $\mathcal{D}$ in the absence of a Hamiltonian.  The first section briefly introduces relevant formalism, the second section discusses the structure of the Liovillians considered in the manuscript in the language of this formalism, and the final section contains the proof.

\subsection{Parametrization of density matrices and superoperators}

An arbitrary density matrix of $N$ spin-$\frac{1}{2}$'s can be expanded as
\begin{align}
\label{supeq:dm_expand}
\hat\rho=\sum_{\{\mu_1,\dots,\mu_N\}}\rho_{\mu_1,\dots \mu_N}\,\hat{\sigma}_1^{\,\mu_1}\otimes\dots\otimes\hat{\sigma}_N^{\,\mu_N};~~~~~\mu_j\in\{1,x,y,z\},
\end{align}
where the coefficients $\rho_{\mu_1,\dots,\mu_N}$ are real numbers and $\hat{\sigma}^1_j=\hat{\mathds{1}}_j$ is the identity operator on site $j$.  For convenience we also introduce a more compact notation in which the set of indices $\{\mu_{1}\dots\mu_N\}$ is replaced by a single collective index $\mu$ (which runs over all possible assignments of values $1,x,y,z$ to all $N$ sites), the basis element $\hat{\sigma}_1^{\,\mu_1}\otimes\dots\otimes\hat{\sigma}_N^{\,\mu_N}$ is denoted simply by $\hat{\mu}$, and the density operator $\hat{\rho}$ (or any other operator) is denoted with a modified ket notation as $\oket{\rho}$.  In this notation, \eref{supeq:dm_expand} becomes $\oket{\rho}=\sum_{\mu}\rho_{\mu}\oket{\mu}$.  The $\rho_{\mu}$ must be real because the density matrix is Hermitian, and the basis elements all are Hermitian, but it is nevertheless useful to consider complex coefficients.  In this notation a general (not-necessarily Hermitian) operator $\hat{A}$ and its Hermitian conjugate $\hat{A}^{\dagger}$ become, respectively, $\oket{A}=\sum_{\mu}A_{\mu}\oket{\mu}$ and $\obra{A}=\sum_{\mu}A^{*}_{\mu}\obra{\mu}$.  The basis elements $\oket{\mu}$ form a vector space (Liouville space) endowed with the inner product $\obra{A}B\rangle\!\rangle\equiv 2^{-N}{\rm Tr}(\hat{A}^{\dagger}\hat{B})=2^{-N}\sum_{\mu\nu}A^{*}_{\mu}B_{\nu}{\rm Tr}(\hat{\mu}\hat{\nu})=\sum_{\mu}A^{*}_{\mu}B_{\mu}$, where we have used the fact that $\hat{\mu}^{\dagger}=\hat{\mu}$ and ${\rm Tr}(\hat{\mu}\hat{\nu})=2^{N}\delta_{\mu,\nu}$. We denote the action of a superoperator on an operator, $\hat{B}=\mathcal{S}(\hat{A})$, with the notation $\oket{B}=\mathcal{S}\oket{A}$.  The superoperator $\mathcal{S}$ inherets a matrix representation from the equation $\oket{B}=\mathcal{S}\oket{A}$ by expanding $\oket{A}=\sum_{\nu}A_{\nu}\oket{\nu}$ and left multiplying both sides of the equation by $\obra{\mu}$, giving $B_{\mu}=\sum_{\nu}\mathcal{S}_{\mu\nu}A_{\nu}$, where $\mathcal{S}_{\mu\nu}\equiv\obra{\mu}\mathcal{S}\oket{\nu}$. Note that a physical Liouvillian must be hermiticity preserving, which implies that its matrix elements in a basis of Hermitian operators (such as the operators $\hat{\mu}$ defined above) are always real.

\subsection{Structure of the Liouvillian}

For our purposes below, an important property of each basis element $\oket{\mu}$ is the number of operators $\hat{\sigma}^x_j$ and $\hat{\sigma}^y_j$ that appear [see \eref{supeq:dm_expand}], which we denote by $d$.  In literature on nuclear-magnetic resonance  Liouville space can be decomposed into a direct sum of subspaces with fixed $d$, and we denote the basis elements of these subspaces by $\oket{d,\mu_d}$, with the understanding that the index $\mu_d$ enumerates only the basis elements in this subspace.  An arbitrary vector within this subspace is denoted by $\oket{d,v}=\sum_{\mu_d} v_{\mu_d}\oket{d,\mu_d}$.  If we define $\mc{P}_d$ as a superoperator that projects onto the subspace spanned by $\oket{d,\mu_d}$, then we can expand $\mathcal{S}=\sum_{d,d'}\mc{P}_d\mathcal{S}\mc{P}_{d'}\equiv\sum_{d,d'}\mathcal{S}^{dd'}$, which has a matrix representation $\mathcal{S}^{dd'}_{\mu_{d}\nu_{d'}}=\obra{d,\mu_d}\mathcal{S}^{dd'}\oket{d',\nu_{d'}}$.

\

The Liouvillian superoperator can be decomposed as $\mathcal{L}=\mathcal{D}+i\mathcal{H}$, where $\mathcal{H}(\star)=-[\hat{H},\star]$.  Because $\mathcal{D}$ and $i\mathcal{H}$ are each (independently) hermiticity preserving, they both have real-valued matrix representations in the chosen basis.  The dissipators we consider in the manuscript obey the condition ${\rm Tr}\big(\hat{\sigma}^z_j\mathcal{D}(\hat{\sigma}^{\pm}_j)\big)={\rm Tr}\big(\hat{\mathds{1}}_j\mathcal{D}(\hat{\sigma}^{\pm}_j)\big)=0$ (\eref{eq:constraint} of the manuscript), which in the language used here implies that $\mathcal{D}^{dd'}=0$ whenever $d<d'$.  In fact, because the dissipator is constructed of local terms, it is straightforward to see that $\mc{D}^{dd'}$ vanishes unless either $d=d'$ or $d=d'+1$.  It is also readily verified that $\mathcal{H}$ is block diagonal in the $d$-subspace decomposition, and that $\mathcal{H}^{00}=0$ (the latter condition following because $\hat{H}$ commutes with all basis elements in the $d=0$ sector).  Hence the complete Liouvillian has the block-lower-triangular structure (dropping subscripts)
\begin{align}
\label{supeq:L_mat}
\mc{L}=
\left\lgroup\begin{matrix}
  \mc{D}^{00} & 0 & 0 & \cdots & 0 & 0 \\[4pt]
  \mc{D}^{10} &  \mc{D}^{11}+i\mc{H}^{11} & 0 & \cdots & 0 & 0 \\[4pt]
  0 &  \mc{D}^{21} &  \mc{D}^{22}+i\mc{H}^{22} & \cdots & 0 & 0 \\[4pt]
\vdots & \vdots & \vdots & \ddots & \vdots & \vdots \\[4pt]
 0 & 0 & 0 & \cdots &  \mc{D}^{N,N-1} &  \mc{D}^{NN}+i\mc{H}^{NN}
\end{matrix}\right\rgroup.
\end{align}
Note that the block-lower-triangular structure of $\mathcal{L}$ of ensures that the time evolution within the $d=0$ subspace, which completely determines the populations in the $z$-basis, is governed by a closed set of equations.  The dynamics within this sector is therefore equivalent to the dynamics of a classical master equation, a situation which sometimes arises in more general quantum systems whenever it is possible to eliminate the contribution of coherences from the equations of motion for populations \cite{PhysRevLett.116.160401,PhysRevA.90.021603}. The eigenvalues of a block-lower triangular matrix are given by the eigenvalues of the diagonal blocks, and therefore
\begin{align}
\label{supeq:eigs}
{\rm eigs}(\mathcal{L})={\rm eigs}(\mathcal{D}^{00})\cup{\rm eigs}(\mathcal{D}^{11}+i\mathcal{H}^{11})\cup\dots\cup{\rm eigs}(\mathcal{D}^{NN}+i\mathcal{H}^{NN}).
\end{align}
Here it is to be understood that the eigenvalues of a projected operator are computed only with respect to vectors supported on that subspace, i.e. the eigenvalues of $\mathcal{D}^{00}$ are given by the eigenvalues of the matrix $\mathcal{D}^{00}_{\mu_0,\nu_0}$.

\subsection{Dissipative gap}

The dissipators we consider are constructed as sums over single-site dissipators, $\mathcal{D}=\sum_{j}\mathcal{D}_j$, each of which independently obeys the constraints (equivalent to \eref{eq:constraint} in the manuscript)
\begin{align}
\label{supeq:constraint}
{\rm Tr}[\hat{\mathds{1}}_j\mathcal{D}_j(\hat{\sigma}_j^x)]={\rm Tr}[\hat{\mathds{1}}_j\mathcal{D}_j(\hat{\sigma}_j^y)]={\rm Tr}[\hat{\sigma}^z_j\mathcal{D}_j(\hat{\sigma}_j^x)]={\rm Tr}[\hat{\sigma}^z_j\mathcal{D}_j(\hat{\sigma}_j^y)]=0.
\end{align}
Here it is understood that $\mathcal{D}_j$ is an operator in the full Liouville space, but because it only acts nontrivially on site $j$ we will (in a slight abuse of notation) temporarily use the symbol $\mathcal{D}_j$ to represent an operator acting on just the Liouville space associated with site $j$, spanned by vectors $\oket{\mu_j}$; regardless of the interpretation of which space $\mathcal{D}_j$ acts on, its eigenvalues are the same up to degeneracies.  With this interpretation in mind, we can represent $\mathcal{D}_j$ as a $4\times4$ real matrix matrix $\obra{\mu_j}\mathcal{D}_j\oket{\nu_j}$.  Using the shorthand $\mathcal{D}_{\mu\nu}\equiv\obra{\mu_j}\mathcal{D}_j\oket{\nu_j}$ (note the temporary suppression of the index $j$), ordering our basis as $\mu=1,z,x,y$, and taking into account the constraints on $\mathcal{D}_j$ imposed in \eref{supeq:constraint}, we have
\begin{align}
\mathcal{D}_{\mu\nu}=\left\lgroup\begin{matrix}[cc|cc]
0 &0 & 0 & 0 \\
\mathcal{D}_{z1} & \mathcal{D}_{zz}  & 0 & 0 \\  \hline 
\mathcal{D}_{x1} & \mathcal{D}_{xz}  & \mathcal{D}_{xx} & \mathcal{D}_{xy} \\
\mathcal{D}_{y1} & \mathcal{D}_{yz}  & \mathcal{D}_{yx} & \mathcal{D}_{yy}
\end{matrix}\right\rgroup.
\end{align}
In order to prove the existence of a dissipative gap for $\mathcal{L}$, we also make the following two assumptions:
\begin{quote}
(1) Each local dissipator $\mathcal{D}_j$ has a unique steady state and a dissipative gap $\Gamma_j$, and ${\rm min}_j\Gamma_j=\Gamma>0$.  Since all eigenvalues of the dissipator must have non-positive real parts, this means that for any nonzero eigenvalue $\lambda$ of $\mathcal{D}_j$, we must have ${\rm Re}(\lambda)\leq -\Gamma_j$.

(2) In addition to \eref{supeq:constraint}, the local dissipators obey the further constraints:
\begin{align}
{\rm Tr}[\hat{\sigma}_j^y\mathcal{D}_j(\hat{\sigma}_j^x)]={\rm Tr}[\hat{\sigma}_j^x\mathcal{D}_j(\hat{\sigma}_j^y)],~~~~{\rm and}~~~~{\rm Tr}[\hat{\sigma}_j^z\mathcal{D}_j(\hat{\mathds{1}}_j)]=0.
\end{align}
\end{quote}
In the notation just introduced, we can restate these conditions as: (1) The matrix $\mathcal{D}_{\mu\nu}$ has a unique zero eigenvalue and a dissipative gap $\Gamma_j$, and (2) The diagonal blocks of $\mathcal{D}_{\mu\nu}$ are symmetric ($\mathcal{D}_{z1}=0$ and $\mathcal{D}_{yx}=\mathcal{D}_{xy}$). Note that the Hamiltonian in \eref{eq:tfim} of the manuscript, when rotated about the $y$ axis and at $\Delta=0$, becomes of the form in \eref{eq:Hamiltonian} of the manuscript, while the jump operators become $\hat{J}_j=\hat{\sigma}_j^y-i\hat{\sigma}_j^z$.  It is straightforward to verify that conditions (1) and (2) are satisfied for the corresponding dissipator, with $\Gamma=\gamma$.

\

As mentioned above, a block-lower-triangular matrix of this form shares its eigenvalues with the diagonal blocks.  Let's denote the set of eigenvalues of the upper-left block by $\mathscr{E}_{j}^0=\{0,\varepsilon^0\}$, and the set of eigenvalues of the lower-right block by $\mathscr{E}_{j}^1=\{\varepsilon^1_a,\varepsilon^1_b\}$ (it is straightforward to see that the upper-left block must have a zero eigenvalue).  By conditions (1) and (2), we know that the three non-zero eigenvalues are real, and that $\varepsilon^0,\varepsilon^1_a,\varepsilon^1_b\leq-\Gamma_j$.  Because each $\mathcal{D}_j$ is supported on a single site, the eigenvalues of the full dissipator $\mc{D}=\sum_{j}\mathcal{D}_j$ are simply sums of eigenvalues from each $\mathcal{D}_j$.  Eigenvalues of the projected dissipator $\mathcal{D}^{dd}$ can be constructed as follows: Defining $\mathscr{J}_d$ to be a set containing $d$ sites, and also defining the notation $\mathscr{A}+\mathscr{B}=\{a+b:~a\in\mathscr{A},~b\in\mathscr{B}\}$, we have
\begin{align}
{\rm eigs}(\mathcal{D}^{dd}) =\bigcup_{\mathscr{J}_d}\bigg(\sum_{j\in\mathscr{J}_d}\mathscr{E}_j^1+\sum_{j\notin\mathscr{J}_d}\mathscr{E}_j^0\bigg).
\end{align}
We can draw two immediate conclusions:
\begin{quote}
(A) $\mathcal{D}^{00}$ has precisely one zero eigenvalue, obtained when the zero eigenvalue in $\mathscr{E}_j^0$ is chosen for all $j$, and any other eigenvalue $\varepsilon$ satisfies $\varepsilon\leq-\Gamma$.

(B) Any eigenvalue $\varepsilon$ of $\mathcal{D}^{dd}$ (with $d>0$) is real and satisfies $\varepsilon<-d\Gamma$.
\end{quote}
Now consider an arbitrary normalized eigenvector of $\mathcal{L}^{dd}$ in the $d\neq 0$ subspace, $\oket{d,v}$, with eigenvalue $\varepsilon=x+iy$, and write
\begin{align}
\obra{d,v}\mathcal{L}^{dd}\oket{d,v}=\obra{d,v}\mathcal{D}^{dd}\oket{d,v}+i\obra{d,v}\mathcal{H}^{dd}\oket{d,v}=x+iy.
\end{align}
It is straightforward to show that the superoperator $\mathcal{H}^{dd}$ is Hermitian (for any Hamiltonian), and thus $x={\rm Re}(\obra{d,v}\mathcal{D}^{dd}\oket{d,v})$.  In general, there is not a simple relation between the real part of the expectation value of an operator and the real parts of its eigenvalues.  However, assumption (2) guarantees that $\mathcal{D}^{dd}$ is a real symmetric matrix.  Therefore, we have
\begin{align}
\label{supeq:ineq}
x=\obra{d,v}\mathcal{D}^{dd}\oket{d,v}\leq -d\Gamma,
\end{align}
where the inequality follows because the expectation value of a \emph{real symmetric matrix} (more generally a Hermitian matrix) is bounded between its smallest and largest eigenvalues.  Returning to \eref{supeq:eigs}, keeping in mind conclusion (A), and using \eref{supeq:ineq}, we find that $\mathcal{L}$ has a single zero eigenvalue, and a dissipative gap of at least $\Gamma$.

\end{widetext}


\begin{thebibliography}{55}%
\makeatletter
\providecommand \@ifxundefined [1]{%
 \@ifx{#1\undefined}
}%
\providecommand \@ifnum [1]{%
 \ifnum #1\expandafter \@firstoftwo
 \else \expandafter \@secondoftwo
 \fi
}%
\providecommand \@ifx [1]{%
 \ifx #1\expandafter \@firstoftwo
 \else \expandafter \@secondoftwo
 \fi
}%
\providecommand \natexlab [1]{#1}%
\providecommand \enquote  [1]{``#1''}%
\providecommand \bibnamefont  [1]{#1}%
\providecommand \bibfnamefont [1]{#1}%
\providecommand \citenamefont [1]{#1}%
\providecommand \href@noop [0]{\@secondoftwo}%
\providecommand \href [0]{\begingroup \@sanitize@url \@href}%
\providecommand \@href[1]{\@@startlink{#1}\@@href}%
\providecommand \@@href[1]{\endgroup#1\@@endlink}%
\providecommand \@sanitize@url [0]{\catcode `\\12\catcode `\$12\catcode
  `\&12\catcode `\#12\catcode `\^12\catcode `\_12\catcode `\%12\relax}%
\providecommand \@@startlink[1]{}%
\providecommand \@@endlink[0]{}%
\providecommand \url  [0]{\begingroup\@sanitize@url \@url }%
\providecommand \@url [1]{\endgroup\@href {#1}{\urlprefix }}%
\providecommand \urlprefix  [0]{URL }%
\providecommand \Eprint [0]{\href }%
\providecommand \doibase [0]{http://dx.doi.org/}%
\providecommand \selectlanguage [0]{\@gobble}%
\providecommand \bibinfo  [0]{\@secondoftwo}%
\providecommand \bibfield  [0]{\@secondoftwo}%
\providecommand \translation [1]{[#1]}%
\providecommand \BibitemOpen [0]{}%
\providecommand \bibitemStop [0]{}%
\providecommand \bibitemNoStop [0]{.\EOS\space}%
\providecommand \EOS [0]{\spacefactor3000\relax}%
\providecommand \BibitemShut  [1]{\csname bibitem#1\endcsname}%
\let\auto@bib@innerbib\@empty
\bibitem [{\citenamefont {Sondhi}\ \emph {et~al.}(1997)\citenamefont {Sondhi},
  \citenamefont {Girvin}, \citenamefont {Carini},\ and\ \citenamefont
  {Shahar}}]{RevModPhys.69.315}%
  \BibitemOpen
  \bibfield  {author} {\bibinfo {author} {\bibfnamefont {S.~L.}\ \bibnamefont
  {Sondhi}}, \bibinfo {author} {\bibfnamefont {S.~M.}\ \bibnamefont {Girvin}},
  \bibinfo {author} {\bibfnamefont {J.~P.}\ \bibnamefont {Carini}}, \ and\
  \bibinfo {author} {\bibfnamefont {D.}~\bibnamefont {Shahar}},\ }\bibfield
  {title} {\enquote {\bibinfo {title} {Continuous quantum phase transitions},}\
  }\href {\doibase 10.1103/RevModPhys.69.315} {\bibfield  {journal} {\bibinfo
  {journal} {Rev. Mod. Phys.}\ }\textbf {\bibinfo {volume} {69}},\ \bibinfo
  {pages} {315--333} (\bibinfo {year} {1997})}\BibitemShut {NoStop}%
\bibitem [{\citenamefont {Sachdev}(2011)}]{sachdev_book}%
  \BibitemOpen
  \bibfield  {author} {\bibinfo {author} {\bibfnamefont {S.}~\bibnamefont
  {Sachdev}},\ }\href@noop {} {\emph {\bibinfo {title} {Quantum Phase
  Transitions}}}\ (\bibinfo  {publisher} {Cambridge University Press},\
  \bibinfo {year} {2011})\BibitemShut {NoStop}%
\bibitem [{\citenamefont {Calabrese}\ and\ \citenamefont
  {Cardy}(2005)}]{1742.5468.2005.04.P04010}%
  \BibitemOpen
  \bibfield  {author} {\bibinfo {author} {\bibfnamefont {P.}~\bibnamefont
  {Calabrese}}\ and\ \bibinfo {author} {\bibfnamefont {J.}~\bibnamefont
  {Cardy}},\ }\bibfield  {title} {\enquote {\bibinfo {title} {Evolution of
  entanglement entropy in one-dimensional systems},}\ }\href
  {http://stacks.iop.org/1742-5468/2005/i=04/a=P04010} {\bibfield  {journal}
  {\bibinfo  {journal} {J. Stat. Mech.}\ }\textbf {\bibinfo {volume} {2005}},\
  \bibinfo {pages} {P04010} (\bibinfo {year} {2005})}\BibitemShut {NoStop}%
\bibitem [{\citenamefont {Fagotti}\ and\ \citenamefont
  {Calabrese}(2008)}]{PhysRevA.78.010306}%
  \BibitemOpen
  \bibfield  {author} {\bibinfo {author} {\bibfnamefont {M.}~\bibnamefont
  {Fagotti}}\ and\ \bibinfo {author} {\bibfnamefont {P.}~\bibnamefont
  {Calabrese}},\ }\bibfield  {title} {\enquote {\bibinfo {title} {Evolution of
  entanglement entropy following a quantum quench: Analytic results for the
  {$XY$} chain in a transverse magnetic field},}\ }\href {\doibase
  10.1103/PhysRevA.78.010306} {\bibfield  {journal} {\bibinfo  {journal} {Phys.
  Rev. A}\ }\textbf {\bibinfo {volume} {78}},\ \bibinfo {pages} {010306}
  (\bibinfo {year} {2008})}\BibitemShut {NoStop}%
\bibitem [{\citenamefont {Prosen}(2008)}]{1367.2630.10.4.043026}%
  \BibitemOpen
  \bibfield  {author} {\bibinfo {author} {\bibfnamefont {T.}~\bibnamefont
  {Prosen}},\ }\bibfield  {title} {\enquote {\bibinfo {title} {Third
  quantization: A general method to solve master equations for quadratic open
  {F}ermi systems},}\ }\href {http://stacks.iop.org/1367-2630/10/i=4/a=043026}
  {\bibfield  {journal} {\bibinfo  {journal} {New J. Phys.}\ }\textbf {\bibinfo
  {volume} {10}},\ \bibinfo {pages} {043026} (\bibinfo {year}
  {2008})}\BibitemShut {NoStop}%
\bibitem [{\citenamefont {Horstmann}\ \emph {et~al.}(2013)\citenamefont
  {Horstmann}, \citenamefont {Cirac},\ and\ \citenamefont
  {Giedke}}]{PhysRevA.87.012108}%
  \BibitemOpen
  \bibfield  {author} {\bibinfo {author} {\bibfnamefont {B.}~\bibnamefont
  {Horstmann}}, \bibinfo {author} {\bibfnamefont {J.~I.}\ \bibnamefont
  {Cirac}}, \ and\ \bibinfo {author} {\bibfnamefont {G.}~\bibnamefont
  {Giedke}},\ }\bibfield  {title} {\enquote {\bibinfo {title} {Noise-driven
  dynamics and phase transitions in fermionic systems},}\ }\href {\doibase
  10.1103/PhysRevA.87.012108} {\bibfield  {journal} {\bibinfo  {journal} {Phys.
  Rev. A}\ }\textbf {\bibinfo {volume} {87}},\ \bibinfo {pages} {012108}
  (\bibinfo {year} {2013})}\BibitemShut {NoStop}%
\bibitem [{\citenamefont {Huffman}\ \emph {et~al.}(2016)\citenamefont
  {Huffman}, \citenamefont {Banerjee}, \citenamefont {Chandrasekharan},\ and\
  \citenamefont {Wiese}}]{Huffman2016309}%
  \BibitemOpen
  \bibfield  {author} {\bibinfo {author} {\bibfnamefont {E.}~\bibnamefont
  {Huffman}}, \bibinfo {author} {\bibfnamefont {D.}~\bibnamefont {Banerjee}},
  \bibinfo {author} {\bibfnamefont {S.}~\bibnamefont {Chandrasekharan}}, \ and\
  \bibinfo {author} {\bibfnamefont {U.-J.}\ \bibnamefont {Wiese}},\ }\bibfield
  {title} {\enquote {\bibinfo {title} {Real-time evolution of strongly coupled
  fermions driven by dissipation},}\ }\href {\doibase
  http://dx.doi.org/10.1016/j.aop.2016.05.019} {\bibfield  {journal} {\bibinfo
  {journal} {Ann. of Phys.}\ }\textbf {\bibinfo {volume} {372}},\ \bibinfo
  {pages} {309 -- 319} (\bibinfo {year} {2016})}\BibitemShut {NoStop}%
\bibitem [{\citenamefont {{\v Z}nidari{\v c}}(2010)}]{Znidaric:2010if}%
  \BibitemOpen
  \bibfield  {author} {\bibinfo {author} {\bibfnamefont {M.}~\bibnamefont {{\v
  Z}nidari{\v c}}},\ }\bibfield  {title} {\enquote {\bibinfo {title} {Exact
  solution for a diffusive nonequilibrium steady state of an open quantum
  chain},}\ }\href
  {http://iopscience.iop.org/article/10.1088/1742-5468/2010/05/L05002/meta}
  {\bibfield  {journal} {\bibinfo  {journal} {J. Stat. Mech.}\ }\textbf
  {\bibinfo {volume} {2010}},\ \bibinfo {pages} {L05002} (\bibinfo {year}
  {2010})}\BibitemShut {NoStop}%
\bibitem [{\citenamefont {\ifmmode \check{Z}\else
  \v{Z}\fi{}nidari\ifmmode~\check{c}\else \v{c}\fi{}}(2011)}]{Znidaric:2011dp}%
  \BibitemOpen
  \bibfield  {author} {\bibinfo {author} {\bibfnamefont {M.}~\bibnamefont
  {\ifmmode \check{Z}\else \v{Z}\fi{}nidari\ifmmode~\check{c}\else
  \v{c}\fi{}}},\ }\bibfield  {title} {\enquote {\bibinfo {title} {Solvable
  quantum nonequilibrium model exhibiting a phase transition and a matrix
  product representation},}\ }\href {\doibase 10.1103/PhysRevE.83.011108}
  {\bibfield  {journal} {\bibinfo  {journal} {Phys. Rev. E}\ }\textbf {\bibinfo
  {volume} {83}},\ \bibinfo {pages} {011108} (\bibinfo {year}
  {2011})}\BibitemShut {NoStop}%
\bibitem [{\citenamefont {Eisler}(2011)}]{Eisler:2011fb}%
  \BibitemOpen
  \bibfield  {author} {\bibinfo {author} {\bibfnamefont {V.}~\bibnamefont
  {Eisler}},\ }\bibfield  {title} {\enquote {\bibinfo {title} {Crossover
  between ballistic and diffusive transport: {T}he quantum exclusion
  process},}\ }\href
  {http://iopscience.iop.org/article/10.1088/1742-5468/2011/06/P06007}
  {\bibfield  {journal} {\bibinfo  {journal} {J. Stat. Mech.}\ }\textbf
  {\bibinfo {volume} {2011}},\ \bibinfo {pages} {P06007} (\bibinfo {year}
  {2011})}\BibitemShut {NoStop}%
\bibitem [{\citenamefont {{\v Z}unkovi{\v c}}(2014)}]{Zunkovic:2014fw}%
  \BibitemOpen
  \bibfield  {author} {\bibinfo {author} {\bibfnamefont {B.}~\bibnamefont {{\v
  Z}unkovi{\v c}}},\ }\bibfield  {title} {\enquote {\bibinfo {title} {Closed
  hierarchy of correlations in {M}arkovian open quantum systems},}\ }\href
  {http://iopscience.iop.org/article/10.1088/1367-2630/16/1/013042} {\bibfield
  {journal} {\bibinfo  {journal} {New J. Phys.}\ }\textbf {\bibinfo {volume}
  {16}},\ \bibinfo {pages} {013042} (\bibinfo {year} {2014})}\BibitemShut
  {NoStop}%
\bibitem [{\citenamefont {Guo}\ and\ \citenamefont
  {Poletti}(2016{\natexlab{a}})}]{Guo_bosons}%
  \BibitemOpen
  \bibfield  {author} {\bibinfo {author} {\bibfnamefont {C.}~\bibnamefont
  {Guo}}\ and\ \bibinfo {author} {\bibfnamefont {D.}~\bibnamefont {Poletti}},\
  }\bibfield  {title} {\enquote {\bibinfo {title} {Solutions for dissipative
  quadratic open systems: {P}art {I} - bosons},}\ }\href
  {https://arxiv.org/abs/1609.07249} {\bibfield  {journal} {\bibinfo  {journal}
  {arXiv:1609.07249}\ } (\bibinfo {year} {2016}{\natexlab{a}})}\BibitemShut
  {NoStop}%
\bibitem [{\citenamefont {Guo}\ and\ \citenamefont
  {Poletti}(2016{\natexlab{b}})}]{Guo_fermions}%
  \BibitemOpen
  \bibfield  {author} {\bibinfo {author} {\bibfnamefont {C.}~\bibnamefont
  {Guo}}\ and\ \bibinfo {author} {\bibfnamefont {D.}~\bibnamefont {Poletti}},\
  }\bibfield  {title} {\enquote {\bibinfo {title} {Solutions for dissipative
  quadratic open systems: {P}art {II} - fermions},}\ }\href
  {https://arxiv.org/abs/1609.07838} {\bibfield  {journal} {\bibinfo  {journal}
  {arXiv:1609.07838}\ } (\bibinfo {year} {2016}{\natexlab{b}})}\BibitemShut
  {NoStop}%
\bibitem [{\citenamefont {Prosen}\ and\ \citenamefont
  {Pi\v{z}orn}(2008)}]{PhysRevLett.101.105701}%
  \BibitemOpen
  \bibfield  {author} {\bibinfo {author} {\bibfnamefont {Toma\v{z}}\
  \bibnamefont {Prosen}}\ and\ \bibinfo {author} {\bibfnamefont {Iztok}\
  \bibnamefont {Pi\v{z}orn}},\ }\bibfield  {title} {\enquote {\bibinfo {title}
  {Quantum phase transition in a far-from-equilibrium steady state of an {$XY$}
  spin chain},}\ }\href {\doibase 10.1103/PhysRevLett.101.105701} {\bibfield
  {journal} {\bibinfo  {journal} {Phys. Rev. Lett.}\ }\textbf {\bibinfo
  {volume} {101}},\ \bibinfo {pages} {105701} (\bibinfo {year}
  {2008})}\BibitemShut {NoStop}%
\bibitem [{\citenamefont
  {Prosen}(2011{\natexlab{a}})}]{PhysRevLett.106.217206}%
  \BibitemOpen
  \bibfield  {author} {\bibinfo {author} {\bibfnamefont {T.}~\bibnamefont
  {Prosen}},\ }\bibfield  {title} {\enquote {\bibinfo {title} {Open {$XXZ$}
  spin chain: {N}onequilibrium steady state and a strict bound on ballistic
  transport},}\ }\href {\doibase 10.1103/PhysRevLett.106.217206} {\bibfield
  {journal} {\bibinfo  {journal} {Phys. Rev. Lett.}\ }\textbf {\bibinfo
  {volume} {106}},\ \bibinfo {pages} {217206} (\bibinfo {year}
  {2011}{\natexlab{a}})}\BibitemShut {NoStop}%
\bibitem [{\citenamefont
  {Prosen}(2011{\natexlab{b}})}]{PhysRevLett.107.137201}%
  \BibitemOpen
  \bibfield  {author} {\bibinfo {author} {\bibfnamefont {T.}~\bibnamefont
  {Prosen}},\ }\bibfield  {title} {\enquote {\bibinfo {title} {Exact
  nonequilibrium steady state of a strongly driven open {$XXZ$} chain},}\
  }\href {\doibase 10.1103/PhysRevLett.107.137201} {\bibfield  {journal}
  {\bibinfo  {journal} {Phys. Rev. Lett.}\ }\textbf {\bibinfo {volume} {107}},\
  \bibinfo {pages} {137201} (\bibinfo {year} {2011}{\natexlab{b}})}\BibitemShut
  {NoStop}%
\bibitem [{\citenamefont {Prosen}(2015)}]{1751-8121-48-37-373001}%
  \BibitemOpen
  \bibfield  {author} {\bibinfo {author} {\bibfnamefont {T.}~\bibnamefont
  {Prosen}},\ }\bibfield  {title} {\enquote {\bibinfo {title} {Matrix product
  solutions of boundary driven quantum chains},}\ }\href
  {http://stacks.iop.org/1751-8121/48/i=37/a=373001} {\bibfield  {journal}
  {\bibinfo  {journal} {J. Phys. A}\ }\textbf {\bibinfo {volume} {48}},\
  \bibinfo {pages} {373001} (\bibinfo {year} {2015})}\BibitemShut {NoStop}%
\bibitem [{\citenamefont {Lee}\ \emph {et~al.}(2013)\citenamefont {Lee},
  \citenamefont {Gopalakrishnan},\ and\ \citenamefont {Lukin}}]{Lee13}%
  \BibitemOpen
  \bibfield  {author} {\bibinfo {author} {\bibfnamefont {T.~E.}\ \bibnamefont
  {Lee}}, \bibinfo {author} {\bibfnamefont {S.}~\bibnamefont {Gopalakrishnan}},
  \ and\ \bibinfo {author} {\bibfnamefont {M.~D.}\ \bibnamefont {Lukin}},\
  }\bibfield  {title} {\enquote {\bibinfo {title} {Unconventional magnetism via
  optical pumping of interacting spin systems},}\ }\href {\doibase
  10.1103/PhysRevLett.110.257204} {\bibfield  {journal} {\bibinfo  {journal}
  {Phys. Rev. Lett.}\ }\textbf {\bibinfo {volume} {110}},\ \bibinfo {pages}
  {257204} (\bibinfo {year} {2013})}\BibitemShut {NoStop}%
\bibitem [{\citenamefont {Sieberer}\ \emph {et~al.}(2016)\citenamefont
  {Sieberer}, \citenamefont {Buchhold},\ and\ \citenamefont
  {Diehl}}]{0034.4885.79.9.096001}%
  \BibitemOpen
  \bibfield  {author} {\bibinfo {author} {\bibfnamefont {L.~M.}\ \bibnamefont
  {Sieberer}}, \bibinfo {author} {\bibfnamefont {M.}~\bibnamefont {Buchhold}},
  \ and\ \bibinfo {author} {\bibfnamefont {S.}~\bibnamefont {Diehl}},\
  }\bibfield  {title} {\enquote {\bibinfo {title} {Keldysh field theory for
  driven open quantum systems},}\ }\href
  {http://stacks.iop.org/0034-4885/79/i=9/a=096001} {\bibfield  {journal}
  {\bibinfo  {journal} {Rep. Prog. Phys.}\ }\textbf {\bibinfo {volume} {79}},\
  \bibinfo {pages} {096001} (\bibinfo {year} {2016})}\BibitemShut {NoStop}%
\bibitem [{\citenamefont {Kraus}\ \emph {et~al.}(2008)\citenamefont {Kraus},
  \citenamefont {B\"uchler}, \citenamefont {Diehl}, \citenamefont {Kantian},
  \citenamefont {Micheli},\ and\ \citenamefont {Zoller}}]{PhysRevA.78.042307}%
  \BibitemOpen
  \bibfield  {author} {\bibinfo {author} {\bibfnamefont {B.}~\bibnamefont
  {Kraus}}, \bibinfo {author} {\bibfnamefont {H.~P.}\ \bibnamefont
  {B\"uchler}}, \bibinfo {author} {\bibfnamefont {S.}~\bibnamefont {Diehl}},
  \bibinfo {author} {\bibfnamefont {A.}~\bibnamefont {Kantian}}, \bibinfo
  {author} {\bibfnamefont {A.}~\bibnamefont {Micheli}}, \ and\ \bibinfo
  {author} {\bibfnamefont {P.}~\bibnamefont {Zoller}},\ }\bibfield  {title}
  {\enquote {\bibinfo {title} {Preparation of entangled states by quantum
  markov processes},}\ }\href {\doibase 10.1103/PhysRevA.78.042307} {\bibfield
  {journal} {\bibinfo  {journal} {Phys. Rev. A}\ }\textbf {\bibinfo {volume}
  {78}},\ \bibinfo {pages} {042307} (\bibinfo {year} {2008})}\BibitemShut
  {NoStop}%
\bibitem [{\citenamefont {Diehl}\ \emph {et~al.}(2008)\citenamefont {Diehl},
  \citenamefont {Micheli}, \citenamefont {Kantian}, \citenamefont {Kraus},
  \citenamefont {Buchler},\ and\ \citenamefont {Zoller}}]{nphys1073}%
  \BibitemOpen
  \bibfield  {author} {\bibinfo {author} {\bibfnamefont {S.}~\bibnamefont
  {Diehl}}, \bibinfo {author} {\bibfnamefont {A.}~\bibnamefont {Micheli}},
  \bibinfo {author} {\bibfnamefont {A.}~\bibnamefont {Kantian}}, \bibinfo
  {author} {\bibfnamefont {B.}~\bibnamefont {Kraus}}, \bibinfo {author}
  {\bibfnamefont {H.~P.}\ \bibnamefont {Buchler}}, \ and\ \bibinfo {author}
  {\bibfnamefont {P.}~\bibnamefont {Zoller}},\ }\bibfield  {title} {\enquote
  {\bibinfo {title} {Quantum states and phases in driven open quantum systems
  with cold atoms},}\ }\href {http://dx.doi.org/10.1038/nphys1073} {\bibfield
  {journal} {\bibinfo  {journal} {Nat. Phys.}\ }\textbf {\bibinfo {volume}
  {4}},\ \bibinfo {pages} {878--883} (\bibinfo {year} {2008})}\BibitemShut
  {NoStop}%
\bibitem [{\citenamefont {Verstraete}\ \emph {et~al.}(2009)\citenamefont
  {Verstraete}, \citenamefont {Wolf},\ and\ \citenamefont {Cirac}}]{nphys1342}%
  \BibitemOpen
  \bibfield  {author} {\bibinfo {author} {\bibfnamefont {F.}~\bibnamefont
  {Verstraete}}, \bibinfo {author} {\bibfnamefont {M.~M.}\ \bibnamefont
  {Wolf}}, \ and\ \bibinfo {author} {\bibfnamefont {J.~I.}\ \bibnamefont
  {Cirac}},\ }\bibfield  {title} {\enquote {\bibinfo {title} {Quantum
  computation and quantum-state engineering driven by dissipation},}\ }\href
  {http://dx.doi.org/10.1038/nphys1342} {\bibfield  {journal} {\bibinfo
  {journal} {Nat. Phys.}\ }\textbf {\bibinfo {volume} {5}},\ \bibinfo {pages}
  {633--636} (\bibinfo {year} {2009})}\BibitemShut {NoStop}%
\bibitem [{\citenamefont {Dengis}\ \emph {et~al.}(2014)\citenamefont {Dengis},
  \citenamefont {K\"{o}nig},\ and\ \citenamefont {Pastawski}}]{Pastawski_2014}%
  \BibitemOpen
  \bibfield  {author} {\bibinfo {author} {\bibfnamefont {J.}~\bibnamefont
  {Dengis}}, \bibinfo {author} {\bibfnamefont {R.}~\bibnamefont {K\"{o}nig}}, \
  and\ \bibinfo {author} {\bibfnamefont {F.}~\bibnamefont {Pastawski}},\
  }\bibfield  {title} {\enquote {\bibinfo {title} {An optimal dissipative
  encoder for the toric code},}\ }\href
  {http://stacks.iop.org/1367-2630/16/i=1/a=013023} {\bibfield  {journal}
  {\bibinfo  {journal} {New J. Phys.}\ }\textbf {\bibinfo {volume} {16}},\
  \bibinfo {pages} {013023} (\bibinfo {year} {2014})}\BibitemShut {NoStop}%
\bibitem [{\citenamefont {Deng}\ \emph {et~al.}(2002)\citenamefont {Deng},
  \citenamefont {Weihs}, \citenamefont {Santori}, \citenamefont {Bloch},\ and\
  \citenamefont {Yamamoto}}]{Deng199}%
  \BibitemOpen
  \bibfield  {author} {\bibinfo {author} {\bibfnamefont {H.}~\bibnamefont
  {Deng}}, \bibinfo {author} {\bibfnamefont {G.}~\bibnamefont {Weihs}},
  \bibinfo {author} {\bibfnamefont {C.}~\bibnamefont {Santori}}, \bibinfo
  {author} {\bibfnamefont {J.}~\bibnamefont {Bloch}}, \ and\ \bibinfo {author}
  {\bibfnamefont {Y.}~\bibnamefont {Yamamoto}},\ }\bibfield  {title} {\enquote
  {\bibinfo {title} {Condensation of semiconductor microcavity exciton
  polaritons},}\ }\href {\doibase 10.1126/science.1074464} {\bibfield
  {journal} {\bibinfo  {journal} {Science}\ }\textbf {\bibinfo {volume}
  {298}},\ \bibinfo {pages} {199--202} (\bibinfo {year} {2002})}\BibitemShut
  {NoStop}%
\bibitem [{\citenamefont {Kasprzak}\ \emph {et~al.}(2006)\citenamefont
  {Kasprzak}, \citenamefont {Richard}, \citenamefont {Kundermann},
  \citenamefont {Baas}, \citenamefont {Jeambrun}, \citenamefont {Keeling},
  \citenamefont {Marchetti}, \citenamefont {Szymanska}, \citenamefont {Andre},
  \citenamefont {Staehli}, \citenamefont {Savona}, \citenamefont {Littlewood},
  \citenamefont {Deveaud},\ and\ \citenamefont {Dang}}]{Kasprazak2006}%
  \BibitemOpen
  \bibfield  {author} {\bibinfo {author} {\bibfnamefont {J.}~\bibnamefont
  {Kasprzak}}, \bibinfo {author} {\bibfnamefont {M.}~\bibnamefont {Richard}},
  \bibinfo {author} {\bibfnamefont {S.}~\bibnamefont {Kundermann}}, \bibinfo
  {author} {\bibfnamefont {A.}~\bibnamefont {Baas}}, \bibinfo {author}
  {\bibfnamefont {P.}~\bibnamefont {Jeambrun}}, \bibinfo {author}
  {\bibfnamefont {J.~M.~J.}\ \bibnamefont {Keeling}}, \bibinfo {author}
  {\bibfnamefont {F.~M.}\ \bibnamefont {Marchetti}}, \bibinfo {author}
  {\bibfnamefont {M.~H.}\ \bibnamefont {Szymanska}}, \bibinfo {author}
  {\bibfnamefont {R.}~\bibnamefont {Andre}}, \bibinfo {author} {\bibfnamefont
  {J.~L.}\ \bibnamefont {Staehli}}, \bibinfo {author} {\bibfnamefont
  {V.}~\bibnamefont {Savona}}, \bibinfo {author} {\bibfnamefont {P.~B.}\
  \bibnamefont {Littlewood}}, \bibinfo {author} {\bibfnamefont
  {B.}~\bibnamefont {Deveaud}}, \ and\ \bibinfo {author} {\bibfnamefont
  {Le~Si}\ \bibnamefont {Dang}},\ }\bibfield  {title} {\enquote {\bibinfo
  {title} {Bose-einstein condensation of exciton polaritons},}\ }\href
  {http://dx.doi.org/10.1038/nature05131} {\bibfield  {journal} {\bibinfo
  {journal} {Nature}\ }\textbf {\bibinfo {volume} {443}},\ \bibinfo {pages}
  {409--414} (\bibinfo {year} {2006})}\BibitemShut {NoStop}%
\bibitem [{\citenamefont {Szyma\ifmmode~\acute{n}\else \'{n}\fi{}ska}\ \emph
  {et~al.}(2006)\citenamefont {Szyma\ifmmode~\acute{n}\else \'{n}\fi{}ska},
  \citenamefont {Keeling},\ and\ \citenamefont
  {Littlewood}}]{PhysRevLett.96.230602}%
  \BibitemOpen
  \bibfield  {author} {\bibinfo {author} {\bibfnamefont {M.~H.}\ \bibnamefont
  {Szyma\ifmmode~\acute{n}\else \'{n}\fi{}ska}}, \bibinfo {author}
  {\bibfnamefont {J.}~\bibnamefont {Keeling}}, \ and\ \bibinfo {author}
  {\bibfnamefont {P.~B.}\ \bibnamefont {Littlewood}},\ }\bibfield  {title}
  {\enquote {\bibinfo {title} {Nonequilibrium quantum condensation in an
  incoherently pumped dissipative system},}\ }\href {\doibase
  10.1103/PhysRevLett.96.230602} {\bibfield  {journal} {\bibinfo  {journal}
  {Phys. Rev. Lett.}\ }\textbf {\bibinfo {volume} {96}},\ \bibinfo {pages}
  {230602} (\bibinfo {year} {2006})}\BibitemShut {NoStop}%
\bibitem [{\citenamefont {Byrnes}\ \emph {et~al.}(2014)\citenamefont {Byrnes},
  \citenamefont {Kim},\ and\ \citenamefont {Yamamoto}}]{Byrnes2014}%
  \BibitemOpen
  \bibfield  {author} {\bibinfo {author} {\bibfnamefont {T.}~\bibnamefont
  {Byrnes}}, \bibinfo {author} {\bibfnamefont {N.~Y.}\ \bibnamefont {Kim}}, \
  and\ \bibinfo {author} {\bibfnamefont {Y.}~\bibnamefont {Yamamoto}},\
  }\bibfield  {title} {\enquote {\bibinfo {title} {Exciton-polariton
  condensates},}\ }\href {http://dx.doi.org/10.1038/nphys3143} {\bibfield
  {journal} {\bibinfo  {journal} {Nat. Phys.}\ }\textbf {\bibinfo {volume}
  {10}},\ \bibinfo {pages} {803--813} (\bibinfo {year} {2014})}\BibitemShut
  {NoStop}%
\bibitem [{\citenamefont {Rodriguez}\ \emph {et~al.}(2016)\citenamefont
  {Rodriguez}, \citenamefont {Amo}, \citenamefont {Sagnes}, \citenamefont
  {Le~Gratiet}, \citenamefont {Galopin}, \citenamefont {Lema{\^\i}tre},\ and\
  \citenamefont {Bloch}}]{ncomms11887}%
  \BibitemOpen
  \bibfield  {author} {\bibinfo {author} {\bibfnamefont {S.~R.~K.}\
  \bibnamefont {Rodriguez}}, \bibinfo {author} {\bibfnamefont {A.}~\bibnamefont
  {Amo}}, \bibinfo {author} {\bibfnamefont {I.}~\bibnamefont {Sagnes}},
  \bibinfo {author} {\bibfnamefont {L.}~\bibnamefont {Le~Gratiet}}, \bibinfo
  {author} {\bibfnamefont {E.}~\bibnamefont {Galopin}}, \bibinfo {author}
  {\bibfnamefont {A.}~\bibnamefont {Lema{\^\i}tre}}, \ and\ \bibinfo {author}
  {\bibfnamefont {J.}~\bibnamefont {Bloch}},\ }\bibfield  {title} {\enquote
  {\bibinfo {title} {Interaction-induced hopping phase in driven-dissipative
  coupled photonic microcavities},}\ }\href
  {http://dx.doi.org/10.1038/ncomms11887} {\bibfield  {journal} {\bibinfo
  {journal} {Nat. Comm.}\ }\textbf {\bibinfo {volume} {7}},\ \bibinfo {pages}
  {11887} (\bibinfo {year} {2016})}\BibitemShut {NoStop}%
\bibitem [{\citenamefont {Bohnet}\ \emph {et~al.}(2016)\citenamefont {Bohnet},
  \citenamefont {Sawyer}, \citenamefont {Britton}, \citenamefont {Wall},
  \citenamefont {Rey}, \citenamefont {Foss-Feig},\ and\ \citenamefont
  {Bollinger}}]{Bohnet1297}%
  \BibitemOpen
  \bibfield  {author} {\bibinfo {author} {\bibfnamefont {J.~G.}\ \bibnamefont
  {Bohnet}}, \bibinfo {author} {\bibfnamefont {B.~C.}\ \bibnamefont {Sawyer}},
  \bibinfo {author} {\bibfnamefont {J.~W.}\ \bibnamefont {Britton}}, \bibinfo
  {author} {\bibfnamefont {M.~L.}\ \bibnamefont {Wall}}, \bibinfo {author}
  {\bibfnamefont {A.~M.}\ \bibnamefont {Rey}}, \bibinfo {author} {\bibfnamefont
  {M.}~\bibnamefont {Foss-Feig}}, \ and\ \bibinfo {author} {\bibfnamefont
  {J.~J.}\ \bibnamefont {Bollinger}},\ }\bibfield  {title} {\enquote {\bibinfo
  {title} {Quantum spin dynamics and entanglement generation with hundreds of
  trapped ions},}\ }\href {\doibase 10.1126/science.aad9958} {\bibfield
  {journal} {\bibinfo  {journal} {Science}\ }\textbf {\bibinfo {volume}
  {352}},\ \bibinfo {pages} {1297--1301} (\bibinfo {year} {2016})}\BibitemShut
  {NoStop}%
\bibitem [{\citenamefont {Schindler}\ \emph {et~al.}(2013)\citenamefont
  {Schindler}, \citenamefont {Muller}, \citenamefont {Nigg}, \citenamefont
  {Barreiro}, \citenamefont {Martinez}, \citenamefont {Hennrich}, \citenamefont
  {Monz}, \citenamefont {Diehl}, \citenamefont {Zoller},\ and\ \citenamefont
  {Blatt}}]{10.1038/nphys2630}%
  \BibitemOpen
  \bibfield  {author} {\bibinfo {author} {\bibfnamefont {P.}~\bibnamefont
  {Schindler}}, \bibinfo {author} {\bibfnamefont {M.}~\bibnamefont {Muller}},
  \bibinfo {author} {\bibfnamefont {D.}~\bibnamefont {Nigg}}, \bibinfo {author}
  {\bibfnamefont {J.~T.}\ \bibnamefont {Barreiro}}, \bibinfo {author}
  {\bibfnamefont {E.~A.}\ \bibnamefont {Martinez}}, \bibinfo {author}
  {\bibfnamefont {M.}~\bibnamefont {Hennrich}}, \bibinfo {author}
  {\bibfnamefont {T.}~\bibnamefont {Monz}}, \bibinfo {author} {\bibfnamefont
  {S.}~\bibnamefont {Diehl}}, \bibinfo {author} {\bibfnamefont
  {P.}~\bibnamefont {Zoller}}, \ and\ \bibinfo {author} {\bibfnamefont
  {R.}~\bibnamefont {Blatt}},\ }\bibfield  {title} {\enquote {\bibinfo {title}
  {Quantum simulation of dynamical maps with trapped ions},}\ }\href
  {http://dx.doi.org/10.1038/nphys2630} {\bibfield  {journal} {\bibinfo
  {journal} {Nat. Phys.}\ }\textbf {\bibinfo {volume} {9}},\ \bibinfo {pages}
  {361--367} (\bibinfo {year} {2013})}\BibitemShut {NoStop}%
\bibitem [{\citenamefont {Peyronel}\ \emph {et~al.}(2012)\citenamefont
  {Peyronel}, \citenamefont {Firstenberg}, \citenamefont {Liang}, \citenamefont
  {Hofferberth}, \citenamefont {Gorshkov}, \citenamefont {Pohl}, \citenamefont
  {Lukin},\ and\ \citenamefont {Vuletic}}]{10.1038/nature11361}%
  \BibitemOpen
  \bibfield  {author} {\bibinfo {author} {\bibfnamefont {T.}~\bibnamefont
  {Peyronel}}, \bibinfo {author} {\bibfnamefont {O.}~\bibnamefont
  {Firstenberg}}, \bibinfo {author} {\bibfnamefont {Q.-Y.}\ \bibnamefont
  {Liang}}, \bibinfo {author} {\bibfnamefont {S.}~\bibnamefont {Hofferberth}},
  \bibinfo {author} {\bibfnamefont {A.~V.}\ \bibnamefont {Gorshkov}}, \bibinfo
  {author} {\bibfnamefont {T.}~\bibnamefont {Pohl}}, \bibinfo {author}
  {\bibfnamefont {M.~D.}\ \bibnamefont {Lukin}}, \ and\ \bibinfo {author}
  {\bibfnamefont {V.}~\bibnamefont {Vuletic}},\ }\bibfield  {title} {\enquote
  {\bibinfo {title} {Quantum nonlinear optics with single photons enabled by
  strongly interacting atoms},}\ }\href {http://dx.doi.org/10.1038/nature11361}
  {\bibfield  {journal} {\bibinfo  {journal} {Nature}\ }\textbf {\bibinfo
  {volume} {488}},\ \bibinfo {pages} {57--60} (\bibinfo {year}
  {2012})}\BibitemShut {NoStop}%
\bibitem [{\citenamefont {Carr}\ \emph {et~al.}(2013)\citenamefont {Carr},
  \citenamefont {Ritter}, \citenamefont {Wade}, \citenamefont {Adams},\ and\
  \citenamefont {Weatherill}}]{Carr13}%
  \BibitemOpen
  \bibfield  {author} {\bibinfo {author} {\bibfnamefont {C.}~\bibnamefont
  {Carr}}, \bibinfo {author} {\bibfnamefont {R.}~\bibnamefont {Ritter}},
  \bibinfo {author} {\bibfnamefont {C.~G.}\ \bibnamefont {Wade}}, \bibinfo
  {author} {\bibfnamefont {C.~S.}\ \bibnamefont {Adams}}, \ and\ \bibinfo
  {author} {\bibfnamefont {K.~J.}\ \bibnamefont {Weatherill}},\ }\bibfield
  {title} {\enquote {\bibinfo {title} {Nonequilibrium phase transition in a
  dilute rydberg ensemble},}\ }\href {\doibase 10.1103/PhysRevLett.111.113901}
  {\bibfield  {journal} {\bibinfo  {journal} {Phys. Rev. Lett.}\ }\textbf
  {\bibinfo {volume} {111}},\ \bibinfo {pages} {113901} (\bibinfo {year}
  {2013})}\BibitemShut {NoStop}%
\bibitem [{\citenamefont {Malossi}\ \emph {et~al.}(2014)\citenamefont
  {Malossi}, \citenamefont {Valado}, \citenamefont {Scotto}, \citenamefont
  {Huillery}, \citenamefont {Pillet}, \citenamefont {Ciampini}, \citenamefont
  {Arimondo},\ and\ \citenamefont {Morsch}}]{PhysRevLett.113.023006}%
  \BibitemOpen
  \bibfield  {author} {\bibinfo {author} {\bibfnamefont {N.}~\bibnamefont
  {Malossi}}, \bibinfo {author} {\bibfnamefont {M.~M.}\ \bibnamefont {Valado}},
  \bibinfo {author} {\bibfnamefont {S.}~\bibnamefont {Scotto}}, \bibinfo
  {author} {\bibfnamefont {P.}~\bibnamefont {Huillery}}, \bibinfo {author}
  {\bibfnamefont {P.}~\bibnamefont {Pillet}}, \bibinfo {author} {\bibfnamefont
  {D.}~\bibnamefont {Ciampini}}, \bibinfo {author} {\bibfnamefont
  {E.}~\bibnamefont {Arimondo}}, \ and\ \bibinfo {author} {\bibfnamefont
  {O.}~\bibnamefont {Morsch}},\ }\bibfield  {title} {\enquote {\bibinfo {title}
  {Full counting statistics and phase diagram of a dissipative rydberg gas},}\
  }\href {\doibase 10.1103/PhysRevLett.113.023006} {\bibfield  {journal}
  {\bibinfo  {journal} {Phys. Rev. Lett.}\ }\textbf {\bibinfo {volume} {113}},\
  \bibinfo {pages} {023006} (\bibinfo {year} {2014})}\BibitemShut {NoStop}%
\bibitem [{\citenamefont {Maghrebi}\ and\ \citenamefont
  {Gorshkov}(2016)}]{PhysRevB.93.014307}%
  \BibitemOpen
  \bibfield  {author} {\bibinfo {author} {\bibfnamefont {M.~F.}\ \bibnamefont
  {Maghrebi}}\ and\ \bibinfo {author} {\bibfnamefont {A.~V.}\ \bibnamefont
  {Gorshkov}},\ }\bibfield  {title} {\enquote {\bibinfo {title} {Nonequilibrium
  many-body steady states via keldysh formalism},}\ }\href {\doibase
  10.1103/PhysRevB.93.014307} {\bibfield  {journal} {\bibinfo  {journal} {Phys.
  Rev. B}\ }\textbf {\bibinfo {volume} {93}},\ \bibinfo {pages} {014307}
  (\bibinfo {year} {2016})}\BibitemShut {NoStop}%
\bibitem [{\citenamefont {Overbeck}\ \emph {et~al.}(2016)\citenamefont
  {Overbeck}, \citenamefont {Maghrebi}, \citenamefont {Gorshkov},\ and\
  \citenamefont {Weimer}}]{Weimer2016}%
  \BibitemOpen
  \bibfield  {author} {\bibinfo {author} {\bibfnamefont {V.~R.}\ \bibnamefont
  {Overbeck}}, \bibinfo {author} {\bibfnamefont {M.~F.}\ \bibnamefont
  {Maghrebi}}, \bibinfo {author} {\bibfnamefont {A.~V.}\ \bibnamefont
  {Gorshkov}}, \ and\ \bibinfo {author} {\bibfnamefont {H.}~\bibnamefont
  {Weimer}},\ }\bibfield  {title} {\enquote {\bibinfo {title} {Multicritical
  behavior in dissipative {I}sing models},}\ }\href
  {https://arxiv.org/abs/arXiv:1606.08863} {\bibfield  {journal} {\bibinfo
  {journal} {arXiv:1606.08863}\ } (\bibinfo {year} {2016})}\BibitemShut
  {NoStop}%
\bibitem [{\citenamefont {Weimer}(2015)}]{PhysRevLett.114.040402}%
  \BibitemOpen
  \bibfield  {author} {\bibinfo {author} {\bibfnamefont {H.}~\bibnamefont
  {Weimer}},\ }\bibfield  {title} {\enquote {\bibinfo {title} {Variational
  principle for steady states of dissipative quantum many-body systems},}\
  }\href {\doibase 10.1103/PhysRevLett.114.040402} {\bibfield  {journal}
  {\bibinfo  {journal} {Phys. Rev. Lett.}\ }\textbf {\bibinfo {volume} {114}},\
  \bibinfo {pages} {040402} (\bibinfo {year} {2015})}\BibitemShut {NoStop}%
\bibitem [{\citenamefont {Raussendorf}\ and\ \citenamefont
  {Briegel}(2001)}]{PhysRevLett.86.5188}%
  \BibitemOpen
  \bibfield  {author} {\bibinfo {author} {\bibfnamefont {R.}~\bibnamefont
  {Raussendorf}}\ and\ \bibinfo {author} {\bibfnamefont {H.~J.}\ \bibnamefont
  {Briegel}},\ }\bibfield  {title} {\enquote {\bibinfo {title} {A one-way
  quantum computer},}\ }\href {\doibase 10.1103/PhysRevLett.86.5188} {\bibfield
   {journal} {\bibinfo  {journal} {Phys. Rev. Lett.}\ }\textbf {\bibinfo
  {volume} {86}},\ \bibinfo {pages} {5188--5191} (\bibinfo {year}
  {2001})}\BibitemShut {NoStop}%
\bibitem [{\citenamefont {Kitagawa}\ and\ \citenamefont
  {Ueda}(1993)}]{PhysRevA.47.5138}%
  \BibitemOpen
  \bibfield  {author} {\bibinfo {author} {\bibfnamefont {M.}~\bibnamefont
  {Kitagawa}}\ and\ \bibinfo {author} {\bibfnamefont {M.}~\bibnamefont
  {Ueda}},\ }\bibfield  {title} {\enquote {\bibinfo {title} {Squeezed spin
  states},}\ }\href {\doibase 10.1103/PhysRevA.47.5138} {\bibfield  {journal}
  {\bibinfo  {journal} {Phys. Rev. A}\ }\textbf {\bibinfo {volume} {47}},\
  \bibinfo {pages} {5138--5143} (\bibinfo {year} {1993})}\BibitemShut {NoStop}%
\bibitem [{\citenamefont {M\o{}lmer}\ and\ \citenamefont
  {S\o{}rensen}(1999)}]{PhysRevLett.82.1835}%
  \BibitemOpen
  \bibfield  {author} {\bibinfo {author} {\bibfnamefont {K.}~\bibnamefont
  {M\o{}lmer}}\ and\ \bibinfo {author} {\bibfnamefont {A.}~\bibnamefont
  {S\o{}rensen}},\ }\bibfield  {title} {\enquote {\bibinfo {title}
  {Multiparticle entanglement of hot trapped ions},}\ }\href {\doibase
  10.1103/PhysRevLett.82.1835} {\bibfield  {journal} {\bibinfo  {journal}
  {Phys. Rev. Lett.}\ }\textbf {\bibinfo {volume} {82}},\ \bibinfo {pages}
  {1835--1838} (\bibinfo {year} {1999})}\BibitemShut {NoStop}%
\bibitem [{\citenamefont {Lee}\ \emph {et~al.}(2005)\citenamefont {Lee},
  \citenamefont {Brickman}, \citenamefont {Deslauriers}, \citenamefont
  {Haljan}, \citenamefont {Duan},\ and\ \citenamefont
  {Monroe}}]{1464.4266.7.10.025}%
  \BibitemOpen
  \bibfield  {author} {\bibinfo {author} {\bibfnamefont {P.~J.}\ \bibnamefont
  {Lee}}, \bibinfo {author} {\bibfnamefont {K.-A.}\ \bibnamefont {Brickman}},
  \bibinfo {author} {\bibfnamefont {L.}~\bibnamefont {Deslauriers}}, \bibinfo
  {author} {\bibfnamefont {P.~C.}\ \bibnamefont {Haljan}}, \bibinfo {author}
  {\bibfnamefont {L.-M.}\ \bibnamefont {Duan}}, \ and\ \bibinfo {author}
  {\bibfnamefont {C.}~\bibnamefont {Monroe}},\ }\bibfield  {title} {\enquote
  {\bibinfo {title} {Phase control of trapped ion quantum gates},}\ }\href
  {http://stacks.iop.org/1464-4266/7/i=10/a=025} {\bibfield  {journal}
  {\bibinfo  {journal} {J. Opt. B}\ }\textbf {\bibinfo {volume} {7}},\ \bibinfo
  {pages} {S371} (\bibinfo {year} {2005})}\BibitemShut {NoStop}%
\bibitem [{\citenamefont {Kim}\ \emph {et~al.}(2010)\citenamefont {Kim},
  \citenamefont {Chang}, \citenamefont {Korenblit}, \citenamefont {Islam},
  \citenamefont {Edwards}, \citenamefont {Freericks}, \citenamefont {Lin},
  \citenamefont {Duan},\ and\ \citenamefont {Monroe}}]{10.1038/nature09071}%
  \BibitemOpen
  \bibfield  {author} {\bibinfo {author} {\bibfnamefont {K.}~\bibnamefont
  {Kim}}, \bibinfo {author} {\bibfnamefont {M.~S.}\ \bibnamefont {Chang}},
  \bibinfo {author} {\bibfnamefont {S.}~\bibnamefont {Korenblit}}, \bibinfo
  {author} {\bibfnamefont {R.}~\bibnamefont {Islam}}, \bibinfo {author}
  {\bibfnamefont {E.~E.}\ \bibnamefont {Edwards}}, \bibinfo {author}
  {\bibfnamefont {J.~K.}\ \bibnamefont {Freericks}}, \bibinfo {author}
  {\bibfnamefont {G.~D.}\ \bibnamefont {Lin}}, \bibinfo {author} {\bibfnamefont
  {L.~M.}\ \bibnamefont {Duan}}, \ and\ \bibinfo {author} {\bibfnamefont
  {C.}~\bibnamefont {Monroe}},\ }\bibfield  {title} {\enquote {\bibinfo {title}
  {Quantum simulation of frustrated ising spins with trapped ions},}\ }\href
  {http://dx.doi.org/10.1038/nature09071} {\bibfield  {journal} {\bibinfo
  {journal} {Nature}\ }\textbf {\bibinfo {volume} {465}},\ \bibinfo {pages}
  {590--593} (\bibinfo {year} {2010})}\BibitemShut {NoStop}%
\bibitem [{\citenamefont {Gaebler}\ \emph {et~al.}(2016)\citenamefont
  {Gaebler}, \citenamefont {Tan}, \citenamefont {Lin}, \citenamefont {Wan},
  \citenamefont {Bowler}, \citenamefont {Keith}, \citenamefont {Glancy},
  \citenamefont {Coakley}, \citenamefont {Knill}, \citenamefont {Leibfried},\
  and\ \citenamefont {Wineland}}]{PhysRevLett.117.060505}%
  \BibitemOpen
  \bibfield  {author} {\bibinfo {author} {\bibfnamefont {J.~P.}\ \bibnamefont
  {Gaebler}}, \bibinfo {author} {\bibfnamefont {T.~R.}\ \bibnamefont {Tan}},
  \bibinfo {author} {\bibfnamefont {Y.}~\bibnamefont {Lin}}, \bibinfo {author}
  {\bibfnamefont {Y.}~\bibnamefont {Wan}}, \bibinfo {author} {\bibfnamefont
  {R.}~\bibnamefont {Bowler}}, \bibinfo {author} {\bibfnamefont {A.~C.}\
  \bibnamefont {Keith}}, \bibinfo {author} {\bibfnamefont {S.}~\bibnamefont
  {Glancy}}, \bibinfo {author} {\bibfnamefont {K.}~\bibnamefont {Coakley}},
  \bibinfo {author} {\bibfnamefont {E.}~\bibnamefont {Knill}}, \bibinfo
  {author} {\bibfnamefont {D.}~\bibnamefont {Leibfried}}, \ and\ \bibinfo
  {author} {\bibfnamefont {D.~J.}\ \bibnamefont {Wineland}},\ }\bibfield
  {title} {\enquote {\bibinfo {title} {High-fidelity universal gate set for
  {${^{9}\mathrm{Be}}^{+}$} ion qubits},}\ }\href {\doibase
  10.1103/PhysRevLett.117.060505} {\bibfield  {journal} {\bibinfo  {journal}
  {Phys. Rev. Lett.}\ }\textbf {\bibinfo {volume} {117}},\ \bibinfo {pages}
  {060505} (\bibinfo {year} {2016})}\BibitemShut {NoStop}%
\bibitem [{FN1()}]{FN1}%
  \BibitemOpen
  \href@noop {} {}\bibinfo {note} {Strictly speaking, this statement applies to
  any individual eigenstate and therefore holds for generic equilibrium
  expectation values, but it does not imply that the dynamics of such models is
  classical.}\BibitemShut {Stop}%
\bibitem [{\citenamefont {Kitaev}(2003)}]{Kitaev20032}%
  \BibitemOpen
  \bibfield  {author} {\bibinfo {author} {\bibfnamefont {A.~Y.}\ \bibnamefont
  {Kitaev}},\ }\bibfield  {title} {\enquote {\bibinfo {title} {Fault-tolerant
  quantum computation by anyons},}\ }\href {\doibase
  http://dx.doi.org/10.1016/S0003-4916(02)00018-0"} {\bibfield  {journal}
  {\bibinfo  {journal} {Ann. Phys.}\ }\textbf {\bibinfo {volume} {303}},\
  \bibinfo {pages} {2 -- 30} (\bibinfo {year} {2003})}\BibitemShut {NoStop}%
\bibitem [{FN2()}]{FN2}%
  \BibitemOpen
  \href@noop {} {}\bibinfo {note} {We have assumed that each jump operator is
  supported on a single site. As a result, $\mathcal{D}$ cannot increase the
  support of an operator, and thus we trivially have ${\rm
  Tr}\big(\hat{\sigma}^z_k\mathcal{D}(\hat{\sigma}^{\pm}_j)\big)=0$ for all
  $j\neq k$. Note also that the constraints for $\hat{\sigma}_{j}^{+}$ and
  $\hat{\sigma}_{j}^{-}$ are redundant, as can be seen by taking the Hermitian
  conjugate of either.}\BibitemShut {Stop}%
\bibitem [{\citenamefont {Foss-Feig}\ \emph
  {et~al.}(2013{\natexlab{a}})\citenamefont {Foss-Feig}, \citenamefont
  {Hazzard}, \citenamefont {Bollinger},\ and\ \citenamefont
  {Rey}}]{PhysRevA.87.042101}%
  \BibitemOpen
  \bibfield  {author} {\bibinfo {author} {\bibfnamefont {M.}~\bibnamefont
  {Foss-Feig}}, \bibinfo {author} {\bibfnamefont {K.~R.~A.}\ \bibnamefont
  {Hazzard}}, \bibinfo {author} {\bibfnamefont {J.~J.}\ \bibnamefont
  {Bollinger}}, \ and\ \bibinfo {author} {\bibfnamefont {A.~M.}\ \bibnamefont
  {Rey}},\ }\bibfield  {title} {\enquote {\bibinfo {title} {Nonequilibrium
  dynamics of arbitrary-range ising models with decoherence: An exact analytic
  solution},}\ }\href {\doibase 10.1103/PhysRevA.87.042101} {\bibfield
  {journal} {\bibinfo  {journal} {Phys. Rev. A}\ }\textbf {\bibinfo {volume}
  {87}},\ \bibinfo {pages} {042101} (\bibinfo {year}
  {2013}{\natexlab{a}})}\BibitemShut {NoStop}%
\bibitem [{\citenamefont {Foss-Feig}\ \emph
  {et~al.}(2013{\natexlab{b}})\citenamefont {Foss-Feig}, \citenamefont
  {Hazzard}, \citenamefont {Bollinger}, \citenamefont {Rey},\ and\
  \citenamefont {Clark}}]{1367.2630.15.11.113008}%
  \BibitemOpen
  \bibfield  {author} {\bibinfo {author} {\bibfnamefont {M.}~\bibnamefont
  {Foss-Feig}}, \bibinfo {author} {\bibfnamefont {K.~R.~A.}\ \bibnamefont
  {Hazzard}}, \bibinfo {author} {\bibfnamefont {J.~J.}\ \bibnamefont
  {Bollinger}}, \bibinfo {author} {\bibfnamefont {A.~M.}\ \bibnamefont {Rey}},
  \ and\ \bibinfo {author} {\bibfnamefont {C.~W.}\ \bibnamefont {Clark}},\
  }\bibfield  {title} {\enquote {\bibinfo {title} {Dynamical quantum
  correlations of ising models on an arbitrary lattice and their resilience to
  decoherence},}\ }\href {http://stacks.iop.org/1367-2630/15/i=11/a=113008}
  {\bibfield  {journal} {\bibinfo  {journal} {New J. Phys.}\ }\textbf {\bibinfo
  {volume} {15}},\ \bibinfo {pages} {113008} (\bibinfo {year}
  {2013}{\natexlab{b}})}\BibitemShut {NoStop}%
\bibitem [{FN3()}]{FN3}%
  \BibitemOpen
  \href@noop {} {}\bibinfo {note} {When \eref{eq:constraint} is violated, we
  have verified numerically that it is possible for single-site
  dissipation---together with a finite-range Hamiltonian of the form in
  \eref{eq:Hamiltonian}---to spread correlations beyond twice the range of the
  Hamiltonian.}\BibitemShut {Stop}%
\bibitem [{\citenamefont {Cubitt}\ \emph {et~al.}(2015)\citenamefont {Cubitt},
  \citenamefont {Lucia}, \citenamefont {Michalakis},\ and\ \citenamefont
  {Perez-Garcia}}]{Cubitt2015}%
  \BibitemOpen
  \bibfield  {author} {\bibinfo {author} {\bibfnamefont {T.~S.}\ \bibnamefont
  {Cubitt}}, \bibinfo {author} {\bibfnamefont {A.}~\bibnamefont {Lucia}},
  \bibinfo {author} {\bibfnamefont {S.}~\bibnamefont {Michalakis}}, \ and\
  \bibinfo {author} {\bibfnamefont {D.}~\bibnamefont {Perez-Garcia}},\
  }\bibfield  {title} {\enquote {\bibinfo {title} {Stability of local quantum
  dissipative systems},}\ }\href {\doibase 10.1007/s00220-015-2355-3}
  {\bibfield  {journal} {\bibinfo  {journal} {Comm. Math. Phys.}\ }\textbf
  {\bibinfo {volume} {337}},\ \bibinfo {pages} {1275--1315} (\bibinfo {year}
  {2015})}\BibitemShut {NoStop}%
\bibitem [{\citenamefont {Gisin}\ and\ \citenamefont
  {Percival}(1993)}]{Gisin1992}%
  \BibitemOpen
  \bibfield  {author} {\bibinfo {author} {\bibfnamefont {N.}~\bibnamefont
  {Gisin}}\ and\ \bibinfo {author} {\bibfnamefont {I.~C.}\ \bibnamefont
  {Percival}},\ }\bibfield  {title} {\enquote {\bibinfo {title} {The quantum
  state diffusion picture of physical processes},}\ }\href
  {http://stacks.iop.org/0305-4470/26/i=9/a=019} {\bibfield  {journal}
  {\bibinfo  {journal} {J. Phys. A}\ }\textbf {\bibinfo {volume} {26}},\
  \bibinfo {pages} {2245} (\bibinfo {year} {1993})}\BibitemShut {NoStop}%
\bibitem [{\citenamefont {Barchielli}\ and\ \citenamefont
  {Gregoratti}(2009)}]{barchielli_book}%
  \BibitemOpen
  \bibfield  {author} {\bibinfo {author} {\bibfnamefont {A.}~\bibnamefont
  {Barchielli}}\ and\ \bibinfo {author} {\bibfnamefont {M.}~\bibnamefont
  {Gregoratti}},\ }\href@noop {} {\emph {\bibinfo {title} {Quantum Trajectories
  and Measurements in Continuous Time: {T}he Diffusive Case}}}\ (\bibinfo
  {publisher} {Springer-Verlag},\ \bibinfo {year} {2009})\BibitemShut {NoStop}%
\bibitem [{\citenamefont {Budini}(2001)}]{PhysRevA.64.052110}%
  \BibitemOpen
  \bibfield  {author} {\bibinfo {author} {\bibfnamefont {A.~A.}\ \bibnamefont
  {Budini}},\ }\bibfield  {title} {\enquote {\bibinfo {title} {Quantum systems
  subject to the action of classical stochastic fields},}\ }\href {\doibase
  10.1103/PhysRevA.64.052110} {\bibfield  {journal} {\bibinfo  {journal} {Phys.
  Rev. A}\ }\textbf {\bibinfo {volume} {64}},\ \bibinfo {pages} {052110}
  (\bibinfo {year} {2001})}\BibitemShut {NoStop}%
\bibitem [{\citenamefont {Chenu}\ \emph {et~al.}(2016)\citenamefont {Chenu},
  \citenamefont {Beau}, \citenamefont {Cao},\ and\ \citenamefont {del
  Campo}}]{Chenu2016}%
  \BibitemOpen
  \bibfield  {author} {\bibinfo {author} {\bibfnamefont {A.}~\bibnamefont
  {Chenu}}, \bibinfo {author} {\bibfnamefont {M.}~\bibnamefont {Beau}},
  \bibinfo {author} {\bibfnamefont {J.}~\bibnamefont {Cao}}, \ and\ \bibinfo
  {author} {\bibfnamefont {A.}~\bibnamefont {del Campo}},\ }\bibfield  {title}
  {\enquote {\bibinfo {title} {Quantum simulation of many-body decoherence:
  Noise as a resource},}\ }\href {https://arxiv.org/abs/1608.01317} {\bibfield
  {journal} {\bibinfo  {journal} {arXiv:1608.01317}\ } (\bibinfo {year}
  {2016})}\BibitemShut {NoStop}%
\bibitem [{\citenamefont {Fischer}\ \emph {et~al.}(2016)\citenamefont
  {Fischer}, \citenamefont {Maksymenko},\ and\ \citenamefont
  {Altman}}]{PhysRevLett.116.160401}%
  \BibitemOpen
  \bibfield  {author} {\bibinfo {author} {\bibfnamefont {M.~H.}\ \bibnamefont
  {Fischer}}, \bibinfo {author} {\bibfnamefont {M.}~\bibnamefont {Maksymenko}},
  \ and\ \bibinfo {author} {\bibfnamefont {E.}~\bibnamefont {Altman}},\
  }\bibfield  {title} {\enquote {\bibinfo {title} {Dynamics of a
  many-body-localized system coupled to a bath},}\ }\href {\doibase
  10.1103/PhysRevLett.116.160401} {\bibfield  {journal} {\bibinfo  {journal}
  {Phys. Rev. Lett.}\ }\textbf {\bibinfo {volume} {116}},\ \bibinfo {pages}
  {160401} (\bibinfo {year} {2016})}\BibitemShut {NoStop}%
\bibitem [{\citenamefont {Hoening}\ \emph {et~al.}(2014)\citenamefont
  {Hoening}, \citenamefont {Abdussalam}, \citenamefont {Fleischhauer},\ and\
  \citenamefont {Pohl}}]{PhysRevA.90.021603}%
  \BibitemOpen
  \bibfield  {author} {\bibinfo {author} {\bibfnamefont {M.}~\bibnamefont
  {Hoening}}, \bibinfo {author} {\bibfnamefont {W.}~\bibnamefont {Abdussalam}},
  \bibinfo {author} {\bibfnamefont {M.}~\bibnamefont {Fleischhauer}}, \ and\
  \bibinfo {author} {\bibfnamefont {T.}~\bibnamefont {Pohl}},\ }\bibfield
  {title} {\enquote {\bibinfo {title} {Antiferromagnetic long-range order in
  dissipative rydberg lattices},}\ }\href {\doibase 10.1103/PhysRevA.90.021603}
  {\bibfield  {journal} {\bibinfo  {journal} {Phys. Rev. A}\ }\textbf {\bibinfo
  {volume} {90}},\ \bibinfo {pages} {021603} (\bibinfo {year}
  {2014})}\BibitemShut {NoStop}%
\end{thebibliography}
\end{document}